\documentclass[a4paper,10pt,twoside]{cpc-hepnp}
\usepackage{multicol}
\usepackage{graphicx}
\usepackage{booktabs}
\usepackage{amssymb,bm,mathrsfs,bbm,amscd}
\usepackage[tbtags]{amsmath}
\usepackage{lastpage}
\usepackage{array,multirow,graphicx}

\begin{document}

\fancyhead[c]{\small Chinese Physics C~~~Vol. xx, No. x (201x) xxxxxx}
\fancyfoot[C]{\small 010201-\thepage}

\footnotetext[0]{}

\title{ Spectroscopy, Decay properties and Regge trajectories of the $B$ and $B_s$ mesons  }

\author{%
    Virendrasinh Kher $^{1}$,$^{2}$\email{vhkher@gmail.com}%
\quad Nayneshkumar Devlani $^{1}$\email{nayneshdev@gmail.com}%
\quad Ajay Kumar Rai$^{2}$*\email{raiajayk@gmail.com}%
}
\maketitle

\address{%
$^1$ Applied Physics Department, Polytechnic, The M.S. University of Baroda, Vadodara 390002, Gujarat, India \\
$^2$ Department of Applied Physics, Sardar Vallabhbhai National Institute of Technology, Surat 395007, Gujarat, India \\
}

\begin{abstract}

A Gaussian wave function is used for detailed study of mass spectra of the $B$ and $B_S$ mesons using a Cornell potential incorporated with a $\mathcal{O}(1/m)$ correction in the potential energy term and expansion of the kinetic energy term up to a ${{\cal{O}}\left({\bf p}^{10}\right)}$ for relativistic correction of the Hamiltonian. The predicted excited states for the $B$ and $B_s$ mesons are in very good agreement with results obtained by experiment. We assigned the $B_{2}(5747)$ and $B_{s2}(5840)$ as $1^{3}p_{2}$ state, the $B_{1}(5721)$ and $B_{s1}(5830)$ as $1 p_{1}$ state, the $B_{0}(5732)$ as $1^{3}p_{0}$ state,  the $B_{s1}(5850)$ as $1p_{1}^{'}$ state and the $B(5970)$ as $2^{3}S_{1}$ state. We investigate the Regge trajectories in the $(J,M^{2})$ and $(n_r,M^{2})$ planes with their corresponding parameters.  Branching ratio for leptonic and radiative-leptonic decays are estimated for the $B$ and $B_S$ mesons. Our results are in good agreement with experimental observation as well as outcomes of other theoretical models.
\end{abstract}

\begin{keyword}
Potential Model \and  Mass spectrum \and   Decay constant \and  Regge trajectories.

\end{keyword}

\begin{pacs}
12.39.Jh,12.40.Yx,13.20.Gd,13.20.Fc
\end{pacs}

\footnotetext[0]{\hspace*{-3mm}\raisebox{0.3ex}{$\scriptstyle\copyright$}2013
Chinese Physical Society and the Institute of High Energy Physics
of the Chinese Academy of Sciences and the Institute
of Modern Physics of the Chinese Academy of Sciences and IOP Publishing Ltd}%

\begin{multicols}{2}

\section{Introduction}
\label{intro}

Last decade is known for noteworthy experimental progress, to  understand the spectroscopy of meson containing heavy-light quark. In the case of the $B$ and $B_S$ mesons, the ground state as well as first few orbitally excited states, have been well established experimentally. Although other radially and orbitally excited states require further experimental investigation. {\cite{PDGlatest,Lu:2016,Shah2016,Liu2016,Liu2015}}. 

In 2013, the CDF collaboration investigated the $B^{0}\pi^{+}$ and $B^{+}\pi^{-}$ invariant mass distributions using data at $\sqrt{s}=1.96 $ TeV $p\bar{p}$ collisions corresponding to integrated luminosity of 9.6 $ {fb}^{-1}$. They found evidence for a new resonance the $B(5970)$. The reported mass and width of the neutral state are $5978\pm5\pm12$ MeV and $70_{-20}^{+30}\pm30$ MeV respectively. More recently the LHCb collaboration also studied the $B^{0}\pi^{+}$ and $B^{+}\pi^{-}$ invariant mass distribution using $\sqrt{s}=7$ and $\sqrt{s}=8$ TeV corresponding to unified luminosity of 3.0 ${fb}^{-1}$. Precise measurements were made for mass and width of the $B_{1}(5721)$ and $B_{2}^{\star}(5747)$ as well as two excited state the $B_{J}(5840)^{0,+}$ and $B_{J}(5960)^{0,+}$ were also observed \cite{Aaij:2015, chen:2016}. 

With the experimental observation, the states for the $B$ and $B_s$ mesons have already been predicted by various theoretical models. These prediction employed various relativistic or relativized quark model, potential model, Bethe-Salpeter Equation as well as constituent quark model based on Dirac Equation \cite{Lu:2016,Shah2016,Liu2016,Godfrey1985,Colangelo1993,DiPierro2001,Yang2012,Ebert2011jc,Eisner2013}. Experimental exploration of newly observed and unconfirmed states of the $B$ and $B_s$ mesons, motivate us to carry out a comprehensive theoretical study.

In this article, we employ a potential model, incorporating corrections to the potential energy part besides the kinematic relativistic correction of the Hamiltonian to understanding the $B$ and $B_s$ mesons. Using our predicted masses of the $B$ and $B_s$ mesons, we  plot the Regge trajectories both in the $(M^{2}\rightarrow J)$ and $(M^{2}\rightarrow n)$ planes (where, $J$ is the spin and $n$ is the principal quantum number), which play vital role to identify any new (experimentally) excited state as well as for information about quantum numbers of the particular state \cite{Ebert2010}.

In present work, the leptonic and radiative leptonic decay widths are estimated. The radiative leptonic decay of mesons can be equal or larger than the pure-leptonic decay, which can unfold a scope of studying the effect of strong interaction in the decay \cite{Yang:2012}.

The article is organized as follows. In Section {\ref{sec:mass}}, we present the theoretical framework for mass spectra. Section {\ref{sec:lepto}}, introduce the theoretical framework for leptonic, {\bf dileptonic} and radiative leptonic decay widths.{\bf Results of mass spectra, leptonic and radiative-leptonic decays are discussed in Section {\ref{sec:Resu}}}. In Section {\ref{sec:reg}}, we investigate Regge trajectories in the $(J,M^{2})$ and $(n_{r},M^{2})$ planes. Finally, we present the conclusion in Section {\ref{sec:conclusion}}.

\section{Methodology}

\subsection{Cornell potential with ${\cal{O}}\left(\frac{1}{m}\right)$ corrections \label{sec:mass}}

 For spectroscopic study of the $B$ and $B_s$ mesons, we employ the Hamiltonian \cite{Gupta1995,Hwang1997,Kher:2017}. 

\begin{equation}
H=\sqrt{\mathbf{p}^{2}+m_{Q}^{2}}+\sqrt{\mathbf{p}^{2}+m_{\bar{q}}^{2}}+V(\mathbf{r});\label{Eq:hamiltonian}
\end{equation} 
where $\mathbf{p}$ stands for relative momentum of the meson,  $m_{Q}$ stands for mass of heavy quark, $m_{\bar{q}}$ stands for mass of light anti-quark and \ensuremath{V(\mathbf{r})} is the meson potential which can be written as \cite{Koma2006}, 
\begin{equation}
V\left(r\right)=V^{\left(0\right)}\left(r\right)+\left(\frac{1}{m_{Q}}+\frac{1}{m_{\bar{q}}}\right)V^{\left(1\right)}\left(r\right)+{\cal O}\left(\frac{1}{m^{2}}\right)
\end{equation}.
 $V^{\left(0\right)}$ is the Cornell like potential \cite{Eichten1978}, 
 
\begin{equation}
V^{\left(0\right)}(r)=-\frac{\alpha_{c}}{r}+Ar+V_{0}
\end{equation}
$V^{\left(1\right)}\left(r\right)$ is the leading order perturbation theory yields

\begin{equation}
V^{\left(1\right)}\left(r\right)=-C_{F}C_{A}\alpha_{s}^{2}/4r^{2}
\end{equation}
where  $\alpha_{c}=(4/3)\alpha_{S}\left({M^{2}}\right)$;
$\alpha_{S}\left({M^{2}}\right)$ is the strong running coupling constant, $A$ is a potential parameter,  $V_{0}$ is a constant, $C_{F}=4/3$ and $C_{A}=3$ are the Casimir charges \cite{Koma2006}.

Here, we utilize the Ritz variational strategy for the study of the $B$ and $B_s$ mesons. In the heavy-light mesons, the confining interaction plays an important role. We employ a Gaussian wave function, to predict the expectation values of the Hamiltonian \cite{Devlani2013,Devlani2011,Kher:2017}. The Gaussian wave function in the position space has the form 
\begin{eqnarray}
R_{nl}(\mu,r) & = & \mu^{3/2}\left(\frac{2\left(n-1\right)!}{\Gamma\left(n+l+1/2\right)}\right)^{1/2}\left(\mu r\right)^{l}\times\nonumber \\
 &  & e^{-\mu^{2}r^{2}/2}L_{n-1}^{l+1/2}(\mu^{2}r^{2})
\end{eqnarray} and in the momentum space has the form
\begin{eqnarray}
R_{nl}(\mu,p) & = & \frac{\left(-1\right)^{n}}{\mu^{3/2}}\left(\frac{2\left(n-1\right)!}{\Gamma\left(n+l+1/2\right)}\right)^{1/2}\left(\frac{p}{\mu}\right)^{l}\times\nonumber \\
 &  & e^{-{p}^{2}/2\mu^{2}}L_{n-1}^{l+1/2}\left(\frac{p^{2}}{\mu^{2}}\right)
\end{eqnarray}
Here,  $L$ and $\mu$ represents the Laguerre polynomial and the variational parameter respectively. Using the Virial theorem \cite{Hwang1997}, we found the value of variational parameter $\mu$ for each state, for opted value of potential parameter $A$, 
\begin{equation}
\left\langle{K.E.}\right\rangle =\frac{1}{2} \left\langle{\frac{rdV}{dr}}\right\rangle.
\end{equation} 

To justify relativistic approach for quarks within the heavy-light mesons, we expand the kinetic energy of the quarks, retaining powers up to ${{\cal{O}}\left({\bf p}^{10}\right)}$, from the Hamiltonian Equation (\ref{Eq:hamiltonian}) \cite{Kher:2017}. In the Virial theorem, we utilize a momentum space wave function to determine the expectation value of the kinetic energy part, whereas a position space Gaussian wave-function is utilized to determine the expectation value of the potential energy part.

{\bf Here, the center of weight mass is the expectation value of the Hamiltonian. By fixing the potential constant $V_0$, $\alpha_s$ and $A$, we fitted the ground state center of weight mass and matched with the PDG value. Fitted potential parameters are listed in Table\ref{tab:potpar}. Using the following Equation, we fitted the ground state center of weight mass \cite{Rai2008,Rai2002}.}
\begin{equation}
M_{SA}=M_{P}+\frac{3}{4}(M_{V}-M_{P});\label{Eq:rai1-1}
\end{equation}
where $M_{V}$ is vector and $M_{P}$ is pseudoscaler meson ground state mass. Using potential parameter listed in Table (\ref{tab:potpar}), we predicted the $S$, $P$, and $D$ state wave center of weight masses of mesons, which are tabulated in Table (\ref{tab:swavespinB}). For the $nJ$ state comparison, we compute the center of weight mass from the  respective theoretical values as \cite{Rai2008}.
\begin{equation}
M_{CW,n}=\frac{\Sigma_{J}(2J+1)M_{nJ}}{\Sigma_{J}(2J+1)}\label{Eq:rai2-1}
\end{equation}
where, $M_{CW,n}$ is the center of weight mass of the $n$ state and $M_{nJ}$ is the meson mass in the $nJ$ state. 
The hyperfine and spin-orbit shifting of the low-lying $S$, $P$ and $D$ state have been estimated by the spin-dependent part of the conventional one gluon exchange potential between the quark and anti-quark \cite{Eichten1994,Gromes1984,Gershtein1995,Kher:2017}
\begin{eqnarray}
V_{SD}(\mathbf{r}) & = & \left(\frac{\mathbf{L\cdot S_{Q}}}{2m_{Q}^{2}}+\frac{\mathbf{L\cdot S_{\bar{q}}}}{2m_{\bar{q}}^{2}}\right)\left(-\frac{dV^{\left(0\right)}(r)}{rdr}+\frac{8}{3}\alpha_{S}\frac{1}{r^{3}}\right)+\nonumber \\
 &  & \frac{4}{3}\alpha_{S}\frac{1}{m_{Q}m_{\bar{q}}}\frac{\mathbf{L\cdot S}}{r^{3}}+\frac{4}{3}\alpha_{S}\frac{2}{3m_{Q}m_{\bar{q}}}\mathbf{S_{Q}\cdot S_{\bar{q}}}4\pi\delta(\mathbf{r})\label{eq:spinhyperfine} \nonumber\\
 &  & +\frac{4}{3}\alpha_{S}\frac{1}{m_{Q}m_{\bar{q}}}\Biggl\{3(\mathbf{S_{Q}\cdot n})(\mathbf{S_{\bar{q}}\cdot n})-\nonumber \\
 &  & (\mathbf{S_{Q}\cdot S_{\bar{q}}})\Biggr\}\frac{1}{r^{3}},\ \quad\mathbf{n}=\frac{\mathbf{r}}{r} 
\end{eqnarray}

where $V^{0}(r)$ stand for the phenomenological potential. In the spin-dependent part, first part stands for the relativistic corrections to the potential $V^{0}(r)$, second part stands for the spin-orbital interaction, third part stands for conventional spin-spin interaction and fourth part stands for tensor interaction.
 
{\bf Mass eigenstates for heavy-light meson are constructed by $jj$ coupling. The quantum numbers \ensuremath{\mathbf{S_{Q}}}  and the light degrees of freedom \ensuremath{\mathbf{j_{\bar{q}}}=\mathbf{s_{\bar{q}}}+\mathbf{L}}, are individually conserved. Here, $\mathbf{S}_{\mathbf{Q}}$ is the heavy quark spin, \ensuremath{\mathbf{s_{\bar{q}}}}  is the light quark spin and \ensuremath{\mathbf{L}}  is the orbital angular momentum of the light quark. The quantum numbers of the excited \ensuremath{\mathbf{L}=1}  states are formed by combining \ensuremath{\mathbf{S_{Q}}}  and \ensuremath{\mathbf{j_{\bar{q}}}} . For \ensuremath{\mathbf{L}=1} we have \ensuremath{\mathbf{j_{\bar{q}}}=1/2}  \ensuremath{(\mathbf{J}=0,1)}  and \ensuremath{\mathbf{j_{\bar{q}}}=3/2}  \ensuremath{(\mathbf{J}=1,2)}  states. These states are denoted as $^{3}P_{0}$, \ensuremath{^{1}P_{1}^{\prime}} ( \ensuremath{\mathbf{j_{\bar{q}}}=1/2}), \ensuremath{^{1}P_{1}}  \ensuremath{(\mathbf{j_{\bar{q}}}=3/2)}
 and \ensuremath{^{3}P_{2}}  in the case of the $B$ and $B_s$ meson. \cite{Gershtein1995,Kher:2017}

Independently of the total spin $J$ projection, one has
\begin{eqnarray}
\left|^{2L+1}L_{L+1}\right\rangle  & = & \left|J=L+1,S=1\right\rangle 
\end{eqnarray}
\begin{eqnarray}
\left|^{2L+1}L_{L}\right\rangle  & = & \sqrt{\frac{L}{L+1}}\left|J=L,S=1\right\rangle +\nonumber\\
 &  & \sqrt{\frac{L+1}{2L+1}}\left|J=L,S=0\right\rangle 
\end{eqnarray}
\begin{eqnarray}
\left|^{2L-1}L_{L}\right\rangle  & = & \sqrt{\frac{L+1}{2L+1}}\left|J=L,S=1\right\rangle -\nonumber \\
 &  & \sqrt{\frac{L}{2L+1}}\left|J=L,S=0\right\rangle 
\end{eqnarray}

\noindent where $\left|J,S\right\rangle $ are the state vectors with the given values of the total quark spin  \textbf{$\mathbf{S=s_{\bar{q}}+S_{Q}}$}, hence the potential terms of the order of $1/m_{\bar{q}}m_{Q}$, $1/m_{Q}^{2}$, lead to the mixing of the levels with the different $j_{\bar{q}}$ values at the given $J$ values. The tensor forces (fourth term in equation (\ref{eq:spinhyperfine})) are zero at $L=0$ or $S=0$.

The heavy-heavy flavored meson states with $J=L$ are mixtures of spin-triplet $\left|^{3}L_{L}\right>$ and spin-singlet $\left|^{1}L_{L}\right>$
states: $J=L=1,\ 2,\ 3,\ldots$
\begin{eqnarray}
\left|\psi_{J}\right> & = & \left|^{1}L_{L}\right>\cos{\phi}+\left|^{3}L_{L}\right>\sin{\phi}
\end{eqnarray}
\begin{eqnarray}
\left|\psi_{J}^{\prime}\right> & = & -\left|^{1}L_{L}\right>\sin{\phi}+\left|^{3}L_{L}\right>\cos{\phi} \end{eqnarray}

\noindent where $\phi$ is the mixing angle and the primed state has the heavier mass. Such mixing occurs due to the nondiagonal spin-orbit and tensor terms in Equation (\ref{eq:spinhyperfine}). The masses of the physical states were obtained by diagonalizing the mixing matrix obtained using equation (\ref{eq:spinhyperfine}) \cite{Gershtein1995}.  Charge conjugating $q\overline{b}$ into $b\overline{q}$ flips the sign of the angle and the phase convention depends on the order of coupling $\ensuremath{\mathbf{L}}$, $\ensuremath{\mathbf{S_{Q}}} $ and $\mathbf{s_{\bar{q}}}$. Radiative transitions are particularly sensitive to the $^{3}L_{L}-{}^{1}L_{L}$ mixing angle with predictions giving radically different results in some cases of different models \cite{Godfrey:2016,Barnes:2002}. The values of mixing angles for P and D states are tabulated in Table(\ref{tab:mixangle}).}

In presents study, the quark masses are $m_{u/d}=0.46\; GeV$,  $m_{b}=4.530 \;GeV$ and $m_{s}= 0.586\;GeV$, to reproduce the ground state masses of the  $B$ and $B_s$ mesons. 

\subsection{Leptonic, Radiative leptonic and Dileptonic Branching fractions\label{sec:lepto}}

To predict the leptonic branching fractions for the ($1^{1}S_{0}$) $B$  mesons, we employed the formula 
\begin{equation}
BR=\Gamma\times\tau
\end{equation}
where $\Gamma$ (leptonic decay width) is given by \cite{Silverman1988}
\begin{eqnarray}
\Gamma({B}^{+}\rightarrow l^{+}\nu_{l})& = & \frac{G_{F}^{2}}{8\pi}f_{B}^{2}\left|V_{ub}\right|^{2}m_{l}^{2}\times\nonumber \\
& & \left(1-\frac{m_{l}^{2}}{M_{B}^{2}}\right)^{2}M_{B}\label{Eq:branchingB}
\end{eqnarray}

For the calculation of the radiative leptonic decay $B^{-}\rightarrow\gamma l\bar{\nu}\,(l=e,\mu)$
width, we employ the formula\cite{Cai-Dian2003} 
\begin{equation}
\Gamma(B^{-}\rightarrow\gamma l\bar{\nu})=\frac{\alpha G_{F}^{2}\left|V_{bu}\right|^{2}}{2592\pi^{2}}f_{B^{-}}^{2}m_{B^{-}}^{3}\left[x_{u}+x_{b}\right],\label{Eq:radiative}
\end{equation}
where
\begin{equation}
x_{u}=\left(3-\frac{m_{B^{-}}}{m_{u}}\right)^{2}, 
\end{equation} 
and
\begin{equation} 
 x_{b}=\left(3-2\frac{m_{B^{-}}}{m_{b}}\right)^{2}
\end{equation} 
\\
Due to the conservation of charge, the single charge lepton decay as well as decay to two muons, are forbidden at the primary transition, but such type of decay occurs in higher-order transitions. Due to Cabibbo-Kobayashi-Maskawa and helicity suppression, expectation of very small  branching fraction for the $B^{0}\rightarrow{\mu}^{+}{\mu}{-}$ and $B_{s}^{0}\rightarrow{\mu}^{+}{\mu}^{-}$ compared to the dominant $\bar{b}$ to $\bar{c}$ transitions. Hence, one can consider dileptonic decays as rare decays. The decay width for the $B_{s}^{0}$ and $B^{0}$ mesons is given by \cite{Shah2016,Bobeth2013,Bobeth2013a}

\begin{eqnarray}
\Gamma(B_{q}^{0}\rightarrow l^{+}l^{-})=\frac{G_{F}^{2}}{\pi}\left(\frac{\alpha}{4\pi sin^{2}\Theta_{W}}\right)^{2}f_{B_{q}}^{2}m_{l}^{2}m_{B_{q}}\nonumber\\
\times\sqrt{1-4\frac{m_{l}^{2}}{m_{B_{q}}^{2}}}|V_{tq}^{\star}V_{tb}|^{2}|C_{10}|^{2}\label{Eq:rareB}
\end{eqnarray}

The branching ratio for $B_{q}^{0}\rightarrow l^{+}l^{-}$ is 
\begin{equation}
BR\rightarrow\Gamma_{(B_{q}^{0}\rightarrow l^{+}l^{-})}\times\tau_{B_{q}}
\end{equation}
$G_{F}$ is the Fermi coupling constant, $f_{B_{q}}$ is the corresponding decay constant, and $C_{10}$ is the Wilson coefficient given by \cite{Shah2016,Buchalla1993,Buras1998} 
 
\begin{equation}
C_{10}=\eta_{Y}\frac{x_{t}}{8}\left[\frac{x_{t}+2}{x_{t}-1}+\frac{3x_{t}-6}{(x_{t}-1)^{2}}ln\: x_{t}\right]
\end{equation}
where $\eta_{Y}(=1.026)$ is the next-to-leading-order
correction {\cite{Shah2016,Buras1998}}, $\Theta_{W}(\approx28^{0})$ is the weak mixing angle (Weinberg angle)\cite{Lee2015} and $x_{t}=\left(m_{t}/m_{w}\right)^{2}$.\\ 

The decay constants were obtained from the Van-Royen-Weisskopf formula \cite{VanRoyen1967}. Incorporating the first order QCD correction factor, 
\begin{equation}
f_{P/V}^{2}=\frac{12\left|\psi_{P/V}(0)\right|^{2}}{M_{P/V}}\bar{C^{2}}(\alpha_{S})\label{Eq:decayconst}
\end{equation}
where \ensuremath{\bar{C^{2}}(\alpha_{S})}
 is the QCD correction factor given by\cite{Braaten1995} 
\begin{equation}
\bar{C^{2}}(\alpha_{S})=1-\frac{\alpha_{S}}{\pi}\left[2-\frac{m_{Q}-m_{\bar{q}}}{m_{Q}+m_{\bar{q}}}\ln\frac{m_{Q}}{m_{\bar{q}}}\right].\label{Eq:correction}
\end{equation}.
We calculate the leptonic branching fractions using Equation (\ref{Eq:branchingB}), the radiative leptonic branching ratio using Equation (\ref{Eq:radiative}) for the $B$ meson and the dileptonic branching ratio with corresponding decay width using the Equation (\ref{Eq:rareB}) for the $B$ and $B_s$ mesons. We have took $\tau_{B}=1.638$ ps,  $\tau_{B_s}=1.510$ ps \cite{PDGlatest} and the calculated values of the pseudoscaler decay constant $f_{B}=0.146(0.150)GeV$ and $f_{B_{S}}=0.187(0.203)GeV$ with(without) QCD correction using the masses obtained from Tables ({\ref{tab:massesB}} and \ref{tab:massesBs}). \\

\end{multicols}
 
 \begin{table*}
   \begin{centering}
  \caption{Potential parameters\label{tab:potpar}.}
  \begin{tabular}{cccc}
  \hline
\noalign{\smallskip}
  Meson & $\alpha_{s}$ & $A\;(GeV^{2})$ & $V_{0}\;(GeV)$ \tabularnewline
    
  \hline
    \noalign{\smallskip}
  $B$ & 0.6675 & 0.118 & -0.00742 \tabularnewline
  \noalign{\smallskip}

    $B_{s}$ & 0.59025 & 0.140 & -0.0108\tabularnewline
  
   \hline
  \end{tabular}
  \par\end{centering}
  \end{table*}

\begin{table}
\caption{\bf Mixing angles $\theta$ for $B$ and $B_{s}$ meson.\label{tab:mixangle}}

\centering{}%
\begin{tabular}{ccccccc}
\hline 
 \noalign{\smallskip}
 Meson & $\theta_{1}^{0}P$ & $\theta_{2}^{0}P$ & $\theta_{3}^{0}P$ & $\theta_{1}^{0}D$ & $\theta_{2}^{0}D$ & $\theta_{3}^{0}D$\tabularnewline
 \noalign{\smallskip}
\hline
 \noalign{\smallskip}
 $B$ & -14.66 & -15.71 & -16.04 & 71.99 & 71.89 & 72.86\tabularnewline
 \addlinespace
 $B_{s}$ & -9.97 & -12.41 & -13.01 & 72.17 & 72.04 & 71.99\tabularnewline
 \noalign{\smallskip}
 
\hline 
\end{tabular}
\end{table}

 \begin{figure}
       \centering 
       \includegraphics[bb=30bp 60bp 750bp 550bp,clip,width=0.72\textwidth]{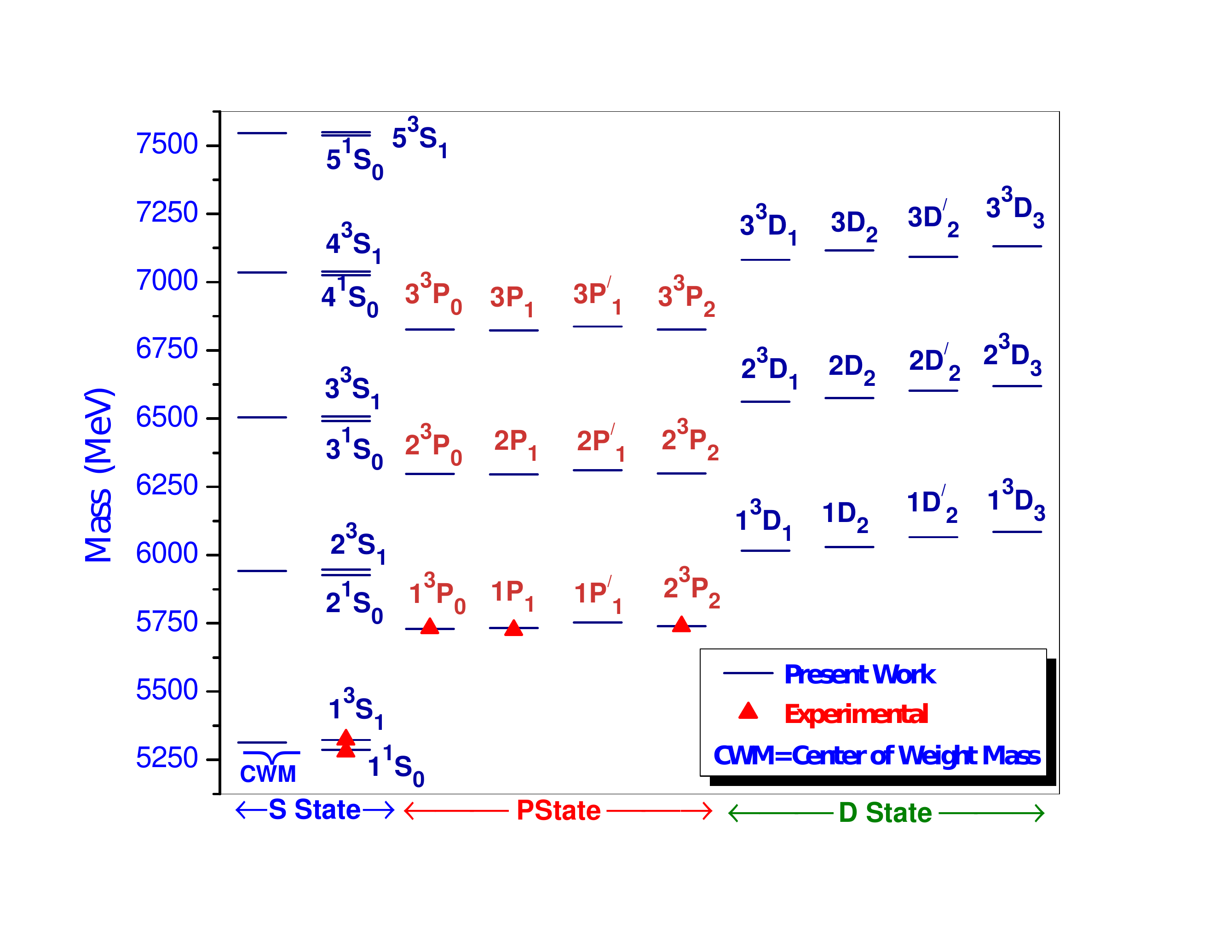}
       \caption{\label{fig:MassB} Mass spectrum of the $B$ meson.}
  \end{figure}
     
  \begin{figure}
       \centering
       \includegraphics[bb=30bp 60bp 750bp 550bp,clip,width=0.72\textwidth]{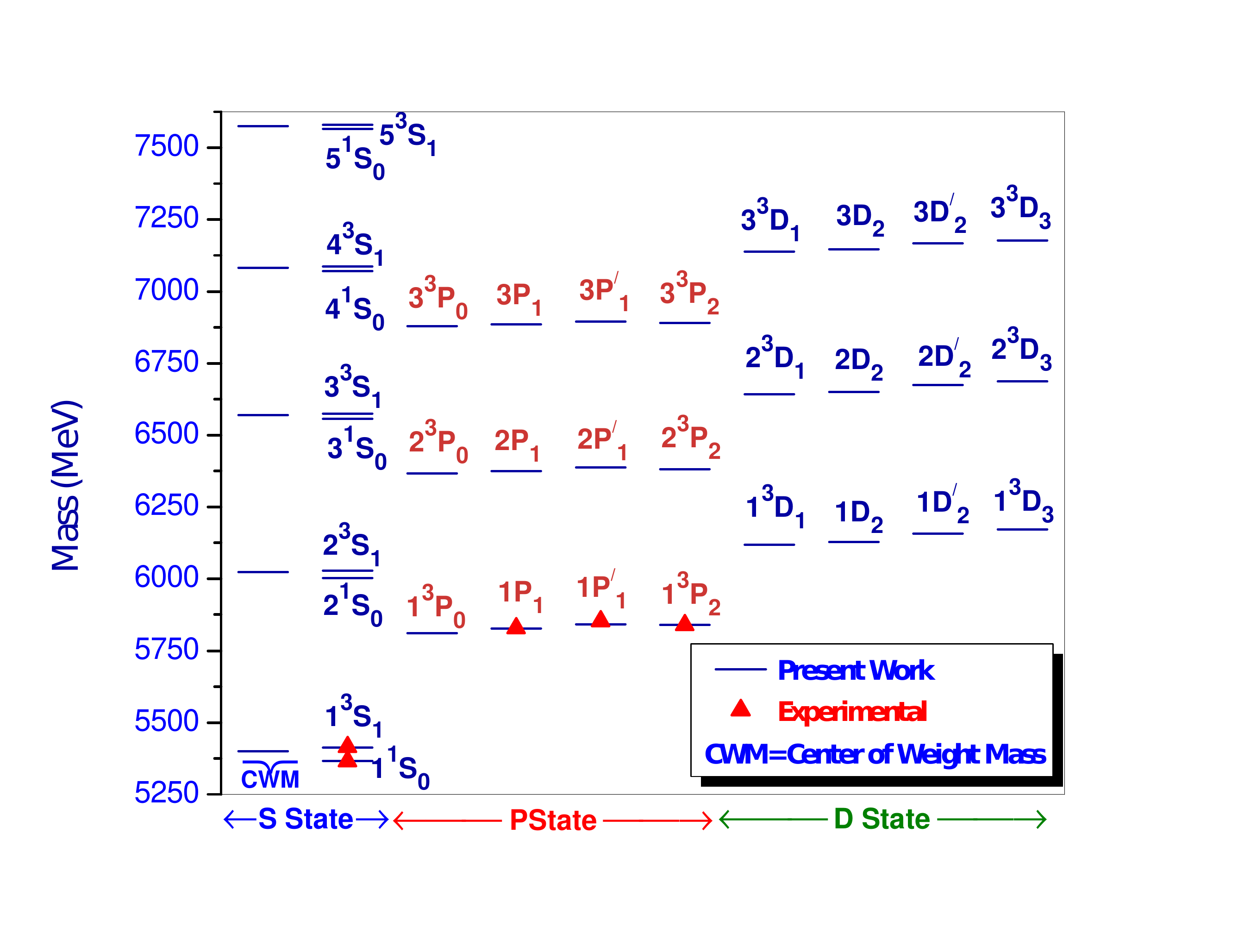}
       \caption{Mass spectrum of the $B_s$ meson.\label{fig:MassBs}}
 \end{figure}

 \begin{table}
 \centering
 \caption{S-P-D-wave center of weight masses (in GeV) of the B meson.\label{tab:swavespinB}}
 
 \begin{tabular}{cccccccccccc}
 \hline 
  \multirow{2}{*}{Meson} &\multirow{2}{*}{State} & \multirow{2}{*}{$\mu$} & \multirow{2}{*}{$M_{CW}$} & \multirow{2}{*}{Expt\cite{PDGlatest}} & \multirow{2}{*}{\cite{Lu:2016}}& \multirow{2}{*}{\cite{Shah2016}} & \multirow{2}{*}{\cite{Liu2016}} &
 \multirow{2}{*}{\cite{Sun:2014}} & \multirow{2}{*}{\cite{Devlani2012}} & \multirow{2}{*}{\cite{Ebert2010}} & \multirow{2}{*}{\cite{Lahde2000}}\tabularnewline
   &  &  &  &  &  &  &  &  & \tabularnewline
 \hline 
  \noalign{\smallskip}
 \multirow{11}{*}{$B$}  & $1S$ & 0.382 & 5.314 & 5.314&5.317 & 5.314 & 5.288  & 5.354  &  5.314 & 5.314 & 5.313\tabularnewline
   & $2S$ & 0.279 & 5.942 &  & 5.932 & 5.819 & 5.903 &5.926 &  5.942 & 5.902 & 5.842\tabularnewline
  &  $3S$ & 0.233 & 6.504 &  & 6.386& 6.251 &  &6.350 & 6.394 & 6.385 & 6.313\tabularnewline
   & $4S$ & 0.206 & 7.036 &  & & 6.647 &  & & 6.778 & 6.785 & 6.347\tabularnewline
  &  $5S$ & 0.187 & 7.546 &  &  & &  &  & & 7.132 & \tabularnewline
     \noalign{\smallskip}
   & $1P$ & 0.309 & 5.740 &  & 5.748& 5.737 & 5.759 & 5.785 & 5.774 & 5.745 & 5.696\tabularnewline
  & $2P$ & 0.249 & 6.301 &  & 6.224& 6.127 & 6.188 & 6.213 & 6.250 & 6.249 & 6.030\tabularnewline
  & $3P$ & 0.216 & 6.828 &  & & 6.482 &  &  & & 6.669 & 6.265\tabularnewline
  \noalign{\smallskip}
 
  & $1D$ & 0.278 & 6.057 &  &6.048 & 6.065 & 6.042 & 6.108 & 6.079 & 6.106 & 5.924\tabularnewline
  &  $2D$ & 0.232 & 6.596 &  &6.467 & 6.429 & 6.377 & 6.466
  & 6.495 & 6.540 & 6.183\tabularnewline
  & $3D$ & 0.205 & 7.110 &  & & 6.769 &  &  &  & & \tabularnewline
  \noalign{\smallskip}
 \hline 
 \noalign{\smallskip}
   
  \multirow{11}{*}{$B_s$} &  $1S$ & 0.470 & 5.401 & 5.401 &5.400 & 5.403 & 5.370 & 5.433 & 5.401 & 5.404 & 5.404\tabularnewline
   & $2S$ & 0.341 & 6.023 &  &5.997 & 5.952 & 5.971 & 6.006 & 6.011 & 5.988 & 5.959\tabularnewline
  &  $3S$ & 0.285 & 6.570 &  &6.430 & 6.425 &  & 6.424 & 6.447 & 6.473 & 6.259\tabularnewline
   &  $4S$ & 0.252 & 7.083 &  & & 6.863 &  & & 6.816 & 6.878 & 6.500\tabularnewline
    & $5S$ & 0.230 & 7.575 &  & & &  &  &  & & \tabularnewline
   \noalign{\smallskip}
    
   &  $1P$ & 0.376 & 5.835 &  &5.827 & 5.838 & 5.838 & 5.858 & 5.851 & 5.844 & 5.805\tabularnewline
   & $2P$ & 0.303 & 6.380 &  & 6.280& 6.233 & 6.254 & 6.290 & 6.310 & 6.343 & 6.161\tabularnewline
   & $3P$ & 0.264 & 6.889 &  & & 6.603 &  &  & & 6.768 & 6.413\tabularnewline
   \noalign{\smallskip}
   
    & $1D$ & 0.337 & 6.150 &  & 6.116& 6.181 & 6.117 & 6.181 & 6.147 & 6.200 & 6.047\tabularnewline
    & $2D$ & 0.283 & 6.668 &  &6.513 & 6.626 & 6.450  & 6.539 & 6.546 & 6.635 & 6.323\tabularnewline
   & $3D$ & 0.251 & 7.162 &  & & 6.912 &  &  &  &  & \tabularnewline
  \hline 
 \end{tabular}
 \end{table}

    \noindent 
    \begin{table*}
    \centering
    \caption{Predicted Masses (in GeV) for the $B$ meson.\label{tab:massesB}}
    
    \centering{}%
    \begin{tabular}{llllllllllll}
    \hline 
    \noalign{\smallskip}
    State &{$J^{P}$} & {Present} & {Expt.\cite{PDGlatest}}& {\cite{Lu:2016}} & {\cite{Shah2016}} & {\cite{Liu2016}} & {\cite{Sun:2014}} & {\cite{Devlani2012}} & {\cite{Ebert2010}} & {\cite{Lahde2000}} \tabularnewline
    &  & work &  &  &  &  & & &  &  \tabularnewline
    \noalign{\smallskip}
    \hline 
    \noalign{\smallskip}
    $1^{1}S_{0}$ & $0^{-}$ & 5.287 & 5.280 $(B^{0})$ &5.280 & 5.279 & 5.273 & 5.309 & 5.266 & 5.280 & 5.277 \tabularnewline
    
    $1^{3}S_{1}$ & $1^{-}$ & 5.323 & 5.325 $(B^{*})$ & 5.329 & 5.325 & 5.331 & 5.369 & 5.330 & 5.326 & 5.325  \tabularnewline
    $2^{1}S_{0}$ & $0^{-}$ & 5.926 & & 5.910 & 5.804 & 5.893 & 5.904& 5.930 & 5.890 & 5.822  \tabularnewline
    $2^{3}S_{1}$ & $1^{-}$ & 5.947 &5.961 $B(5970)$\cite{Aaltonen:2013} & 5.939 & 5.824 & 5.932 & 5.934 & 5.946 & 5.906 & 5.848  \tabularnewline
    $3^{1}S_{0}$ & $0^{-}$ & 6.492 & & 6.369 & 6.242 &  & 6.334 & 6.387 & 6.379 & 6.117  \tabularnewline
    $3^{3}S_{1}$ & $1^{-}$ & 6.508 & & 6.391 & 6.254 &  & 6.355& 6.396 & 6.387 & 6.l36  \tabularnewline
    $4^{1}S_{0}$ & $0^{-}$ & 7.027 & & & 6.641 &  & & 6.773 & 6.781 & 6.335  \tabularnewline
    $4^{3}S_{1}$ & $1^{-}$ & 7.039 & & & 6.649 &  & & 6.779 & 6.786 & 6.351  \tabularnewline
    $5^{1}S_{0}$ & $0^{-}$ & 7.538 & & &  &  &  & & &   \tabularnewline
    $5^{3}S_{1}$ & $1^{-}$ & 7.549 & & &  &  &  & & &   \tabularnewline
    \hline 
    \noalign{\smallskip}
    $1^{3}P_{0}$ & $0^{+}$ & 5.730 & 5.710 $B_{0}(5732)$\cite{Affolder:2001} & 5.683 & 5.697 & 5.740 & 5.756& 5.746 & 5.749 & 5.678  \tabularnewline
    $1P_{1}$ & $1^{+}$ & 5.733 & 5.726 $B_{1}(5721)$ & 5.729& 5.723 & 5.815& 5.782 & 5.764 & 5.723 & 5.686  \tabularnewline
    $1P_{1}^{\prime}$  & $1^{+}$ & 5.752 & & 5.754 & 5.738 & 5.731 &5.779 & 5.785 & 5.774 & 5.699  \tabularnewline
    $1^{3}P_{2}$ & $2^{+}$ & 5.740 & 5.740 $B_{2}(5747)$ & 5.768& 5.754 & 5.746 & 5.796 & 5.779 & 5.741 & 5.704  \tabularnewline
    
    $2^{3}P_{0}$ & $0^{+}$ & 6.297 & & 6.145& 6.053 & 6.188 & 6.214 & 6.225 & 6.221 & 6.010  \tabularnewline
    $2P_{1}$ & $1^{+}$ & 6.295 & & 6.185 & 6.106 & 6.168 & 6.206 & 6.243 & 6.209 & 6.022  \tabularnewline
    $2P_{1}^{\prime}$ & $1^{+}$ & 6.311 & & 6.241& 6.131 & 6.221 & 6.219 & 6.256 & 6.281 & 6.028  \tabularnewline
    $2^{3}P_{2}$ & $2^{+}$ & 6.299 & & 6.253& 6.153 & 6.179 & 6.213 & 6.255 & 6.260 & 6.040  \tabularnewline
    
    $3^{3}P_{0}$ & $0^{+}$ & 6.826 & & & 6.375 &  & & & 6.629 & 6.242  \tabularnewline
    $3P_{1}$ & $1^{+}$ & 6.824 & & & 6.453 &  & & & 6.650 & 6.259  \tabularnewline
    $3P_{1}^{\prime}$ & $1^{+}$ & 6.837 & & & 6.486 &  &  & & 6.685 & 6.260  \tabularnewline
    $3^{3}P_{2}$ & $2^{+}$ & 6.826 & & & 6.518 &  & &  & 6.678 & 6.277 \tabularnewline
    \hline 
    \noalign{\smallskip}
    $1^{3}D_{1}$ & $1^{-}$ & 6.016 & & 6.095& 6.104 & 6.135 & 6.110 & 6.114 & 6.119 & 6.005  \tabularnewline
    $1D_{2}$ & $2^{-}$ & 6.031 & &6.004 & 6.076 & 5.967 & 6.108& 6.125 & 6.121 & 5.955  \tabularnewline
    $1D_{2}^{\prime}$ & $2^{-}$ & 6.065 & & 6.113& 6.065 & 6.152 & 6.113& 6.056 & 6.103 & 5.920  \tabularnewline
    $1^{3}D_{3}$ & $3^{-}$ & 6.085 & & 6.014& 6.041 & 5.976 & 6.105 & 6.060 & 6.091 & 5.871 \tabularnewline
    
    $2^{3}D_{1}$ & $1^{-}$ & 6.562 & & 6.497 & 6.460 & 6.445 & 6.475 & 6.522 & 6.534 & 6.248  \tabularnewline
    $2D_{2}$ & $2^{-}$ & 6.575 & & 6.435 & 6.440 & 6.323 & 6.464& 6.532 & 6.554 & 6.207  \tabularnewline
    $2D_{2}^{\prime}$ & $2^{-}$ & 6.602 & & 6.513 & 6.429 & 6.456 & 6.472 & 6.476 & 6.528 & 6.179  \tabularnewline
    $2^{3}D_{3}$ & $3^{-}$ & 6.619 & & 6.444 & 6.409 & 6.329 & 6.459 & 6.479 & 6.542 & 6.140  \tabularnewline
    
    $3^{3}D_{1}$ & $1^{-}$ & 7.081 & & & 6.795 &  &  & &    \tabularnewline
    $3D_{2}$ & $2^{-}$ & 7.093 & & & 6.768 &  &  &  &  & \tabularnewline
    $3D_{2}^{\prime}$ & $2^{-}$ & 7.116 & & & 6.779 &  &  & & &   \tabularnewline
    $3^{3}D_{3}$ & $3^{-}$ & 7.130 & & & 6.751 &  &  &  & & \tabularnewline
    \hline
    \end{tabular}
    \end{table*}

\begin{table*}
    \centering
  \caption{Predicted Masses (in GeV) for the $B_{S}$ meson.\label{tab:massesBs}}
    \centering{}%
    \begin{tabular}{lllllllllll}
    \hline 
    \noalign{\smallskip}
    State & {$J^{P}$} & Present & {Expt.\cite{PDGlatest}}  & {\cite{Lu:2016}} & {\cite{Shah2016}} & {\cite{Liu2016}} & {\cite{Sun:2014}} & {\cite{Devlani2012}} & {\cite{Ebert2010}} & {\cite{Lahde2000}} \tabularnewline
     &  & work &   & &  &  & & &  &  \tabularnewline
    \hline
    \noalign{\smallskip}
    $1^{1}S_{0}$ & $0^{-}$ & 5.367 & 5.366 $B_{s}^{0}$ & 5.362 & 5.366 & 5.355 & 5.390 & 5.355 & 5.372 & 5.366  \tabularnewline
    $1^{3}S_{1}$ & $1^{-}$ & 5.413 & 5.415 $B_{s}^{*}$ & 5.413 & 5.415 & 5.416 & 5.447 & 5.417 & 5.414 & 5.417 \tabularnewline
    $2^{1}S_{0}$ & $0^{-}$ & 6.003 &  & 5.977 & 5.939 & 5.962 & 5.985 & 5.998 & 5.976 & 5.939 \tabularnewline
    $2^{3}S_{1}$ & $1^{-}$ & 6.029 &  & 6.003 & 5.956 & 5.999 & 6.013 & 6.016 & 5.992 & 5.966  \tabularnewline
    $3^{1}S_{0}$ & $0^{-}$ & 6.556 &  & 6.415 & 6.419 &  & 6.409 & 6.441 & 6.467 & 6.254  \tabularnewline
    $3^{3}S_{1}$ & $1^{-}$ & 6.575 & & 6.435 & 6.427 &  & 6.429 & 6.449 & 6.475 & 6.274  \tabularnewline
    $4^{1}S_{0}$ & $0^{-}$ & 7.071 &  &  & 6.859 &  &  & 6.812 & 6.874 & 6.487  \tabularnewline
    $4^{3}S_{1}$ & $1^{-}$ & 7.087 &  &  & 6.864 &  &  & 6.818 & 6.879 & 6.504  \tabularnewline
    $5^{1}S_{0}$ & $0^{-}$ & 7.565 &  &  &  &  &  &  &  &   \tabularnewline
    $5^{3}S_{1}$ & $1^{-}$ & 7.579 &  &  &  &  &  &  &  &   \tabularnewline
    \hline 
    \noalign{\smallskip}
    $1^{3}P_{0}$ & $0^{+}$ & 5.812 &  & 5.756 & 5.799 & 5.782 & 5.830 & 5.820 & 5.833 & 5.781  \tabularnewline
    $1P_{1}$ & $1^{+}$ & 5.828 & 5.829 $B_{s1}(5830)$ & 5.801 & 5.819 & 5.833 & 5.838 & 5.857 & 5.865 & 5.805  \tabularnewline
    $1P_{1}^{\prime}$ & $1^{+}$ & 5.842 & 5.853 $B_{s1}(5850)$  &5.836 & 5.854 & 5.843 & 5.859 & 5.845 & 5.831 & 5.795 \tabularnewline
    $1^{3}P_{2}$ & $2^{+}$ & 5.840 & 5.840  $B_{s2}(5840)$& 5.851 & 5.849 & 5.848 & 5.875 & 5.859 & 5.842 & 5.815 \tabularnewline
    
    $2^{3}P_{0}$ & $0^{+}$ & 6.367 &  & 6.203 & 6.171 & 6.220 & 6.279 & 6.283 & 6.318 & 6.143  \tabularnewline
    $2P_{1}$ & $1^{+}$ & 6.375 &  & 6.241 & 6.197 & 6.250 & 6.284 &  6.306 & 6.345 & 6.153  \tabularnewline
    $2P_{1}^{\prime}$ & $1^{+}$ & 6.387 &  & 6.297 & 6.278 & 6.256  & 6.291& 6.312 & 6.321 & 6.160  \tabularnewline
    $2^{3}P_{2}$ & $2^{+}$ & 6.382 &  & 6.309 & 6.241 & 6.261 & 6.295& 6.317 & 6.359 & 6.170  \tabularnewline
    
    $3^{3}P_{0}$ & $0^{+}$ & 6.879 &  &  & 6.510 &  & &   & 6.731 & 6.396  \tabularnewline
    $3P_{1}$ & $1^{+}$ & 6.885 &  &  & 6.663 &  &  &  & 6.761 & 6.406  \tabularnewline
    $3P_{1}^{\prime}$ & $1^{+}$ & 6.895 &  &  & 6.543 &  & &   & 6.768 & 6.411  \tabularnewline
    $3^{3}P_{2}$ & $2^{+}$ & 6.890 &  &  & 6.622 &  &  &  & 6.780 & 6.421  \tabularnewline
    \hline 
    \noalign{\smallskip}
    $1^{3}D_{1}$ & $1^{-}$ & 6.119 &  & 6.142 & 6.226 & 6.155 & 6.181 & 6.188 & 6.209 & 6.094  \tabularnewline
    $1D_{2}$ & $2^{-}$ & 6.128 &  & 6.087 & 6.177 & 6.079 & 6.180 &  6.199 & 6.218 & 6.067  \tabularnewline
    $1D_{2}^{\prime}$ & $2^{-}$ & 6.157 &  & 6.159 & 6.209 & 6.172 & 6.185 & 6.110 & 6.189 & 6.043  \tabularnewline
    $1^{3}D_{3}$ & $3^{-}$ & 6.172 &  & 6.096 & 6.145 & 6.088 & 6.178 & 6.188 & 6.191 & 6.016  \tabularnewline
    
    $2^{3}D_{1}$ & $1^{-}$ & 6.642 &  & 6.527 & 6.595 & 6.478 & 6.542 & 6.579 & 6.629 & 6.362  \tabularnewline
    $2D_{2}$ & $2^{-}$ & 6.650 &  & 6.492 & 6.554 & 6.422 & 6.536 &  6.588 & 6.651 & 6.339  \tabularnewline
    $2D_{2}^{\prime}$ & $2^{-}$ & 6.674 &  & 6.542 & 6.585 & 6.490 &6.542 & 6.517 & 6.625 & 6.320 \tabularnewline
    $2^{3}D_{3}$ & $3^{-}$ & 6.687 &  &  6.500 & 6.528 & 6.429 & 6.534 & 6.524 & 6.637 & 6.298  \tabularnewline
    
    $3^{3}D_{1}$ & $1^{-}$ & 7.139 &  & & 6.942 &  &  &  &  &   \tabularnewline
    $3D_{2}$ & $2^{-}$ & 7.147 &  &  & 6.907 &  &  &  &  &   \tabularnewline
    $3D_{2}^{\prime}$ & $2^{-}$ & 7.167 &  &  & 6.936 &  &  &  &  &   \tabularnewline
    $3^{3}D_{3}$ & $3^{-}$ & 7.178 &  &  & 6.885 &  &  &  &  &   \tabularnewline
     
    \hline 
    \end{tabular}
    \end{table*}

  \begin{table}
  \caption{Leptonic branching fractions of the $B$ meson \label{tab:leptobranch}.}
       \centering{}%
   \begin{tabular}{cccc}
          \hline 
  & \multirow{2}{*}{$B^{+}\rightarrow\tau^{+}\nu_{\tau}$} & \multirow{2}{*}{$B^{+}\rightarrow\mu^{+}\nu_{\mu}$} & \multirow{2}{*}{$B^{+}\rightarrow e^{+}\nu_{e}$}\tabularnewline
            
  \multirow{2}{*}{} & \multirow{2}{*}{$BR_{\tau}$} & \multirow{2}{*}{$BR_{\mu}$} & \multirow{2}{*}{$BR_{e}$}\tabularnewline
       \noalign{\smallskip}
       \noalign{\smallskip}
    \cline{2-4}
  \multirow{2}{*}{This work} & \multirow{2}{*}{$0.822\times10^{-4}$} & \multirow{2}{*}{$0.37\times10^{-7}$} & \multirow{2}{*}{$8.64\times10^{-12}$}\tabularnewline
     \noalign{\smallskip} \noalign{\smallskip} \noalign{\smallskip}
     PDG\cite{PDGlatest} & ($1.14\pm0.27)\times10^{-4}$ & $<1.0\times10^{-6}$ & $<9.8\times10^{-7}$\tabularnewline
       \hline
  \end{tabular}
  \end{table}

  \begin{table}
  \begin{centering}
  \caption{ Branching ratio with corresponding rare leptonic decay width of the $B^{0}$ meson. \label{tab:rareleptoB}}
     \begin{tabular}{ccccc}
      \hline
  \noalign{\smallskip}
      \multirow{2}{*}{Process} & \multicolumn{2}{c}{$\Gamma(B_{q}^{0}\rightarrow l^{+}l^{-})(keV)$} & \multicolumn{2}{c}{$BR$}\tabularnewline
             \cline{2-5}
         \noalign{\smallskip}
      & Present & {\cite{Shah2016}} & Present & Others \tabularnewline
              \noalign{\smallskip}
    \hline 
       \noalign{\smallskip}
   $B^{0}\rightarrow\mu^{+}\mu^{-}$ & $4.341\times10^{-17}$ & $4.406\times10^{-17}$ & $1.002\times10^{-10}$ & $\bf \left(3.9_{-1.4}^{+1.6}\right)\times10^{-10}$\cite{PDGlatest}\tabularnewline
  &  &  &  & $<1.1\times10^{-9}$\cite{Chatrchyan2013}\tabularnewline
   &  &  &  & $<9.4\times10^{-10}$\cite{Aaij2012}\tabularnewline
   &  &  &  & $<7.4\times10^{-10}$\cite{Aaij2013}\tabularnewline
   &  &  &  & $1.20\times10^{-10}$\cite{Dimopoulos2011}\tabularnewline
 & & & & $(1.06\pm0.09)\times10^{-10}$\cite{Bobeth2013}\tabularnewline
 &  &  &  & $1.018\times10^{-10}$\cite{Shah2016}\tabularnewline
            \cline{1-5} 
          \noalign{\smallskip}
 $B^{0}\rightarrow\tau^{+}\tau^{-}$ & $9.097\times10^{-15}$ & $9.232\times10^{-15}$ & $2.099\times10^{-8}$ & $<4.1\times10^{-3}$\cite{PDGlatest}\tabularnewline
   &  &  &  & $2.52\times10^{-8}$\cite{Dimopoulos2011}\tabularnewline
   &  &  &  & $(2.22\pm0.19)\times10^{-8}$\cite{Bobeth2013}\tabularnewline
   &  &  &  & $2.133\times10^{-8}$\cite{Shah2016}\tabularnewline
            \cline{1-5} 
           \noalign{\smallskip}
 $B^{0}\rightarrow e^{+}e^{-}$ & $1.016\times10^{-21}$ & $1.028\times10^{-21}$ & $2.345\times10^{-15}$ & $<8.3\times10^{-8}$\cite{PDGlatest}\tabularnewline
    &  &  &  & $2.82\times10^{-15}$\cite{Dimopoulos2011}\tabularnewline
   &  &  &  & $(2.48\pm0.21)\times10^{-15}$\cite{Bobeth2013}\tabularnewline
     &  &  &  & $2.376\times10^{-15}$\cite{Shah2016}\tabularnewline
  \hline 
\end{tabular}
 \par\end{centering}
  \end{table}

\begin{table}
  \begin{centering}
 \caption{ Branching ratio with corresponding rare leptonic decay width of the $B_{s}^{0}$ meson. \label{tab:rareleptoBs}}
  \begin{tabular}{ccccc}
           \hline   \noalign{\smallskip}
  \multirow{2}{*}{Process} & \multicolumn{2}{c}{$\Gamma(B_{q}^{0}\rightarrow l^{+}l^{-})(keV)$} & \multicolumn{2}{c}{$BR$}\tabularnewline
             \cline{2-5}
             \noalign{\smallskip}
    & Present & {\cite{Shah2016}} & Present & Others \tabularnewline
              \noalign{\smallskip}
           \hline
           \noalign{\smallskip}
 $B_{s}^{0}\rightarrow\mu^{+}\mu^{-}$ & $1.101\times10^{-15}$ & $1.583\times10^{-15}$ & $2.529\times10^{-9}$ & $\bf \left(2.9_{-0.6}^{+0.7}\right)\times10^{-9}${\cite{PDGlatest}}\tabularnewline
              &  &  &  & $3.0_{-0.9}^{+1.0}\times10^{-9}${\cite{Chatrchyan2013}}\tabularnewline
             &  &  &  & $3.2_{-1.2}^{+1.5}\times10^{-9}${\cite{Aaij2012}}\tabularnewline
              &  &  &  & $2.9{}_{-1.0}^{+1.1}\times10^{-9}${\cite{Aaij2013}}\tabularnewline
             &  &  &  & $3.40\times10^{-9}${\cite{Dimopoulos2011}}\tabularnewline
             &  &  &  & $(3.65\pm0.23)\times10^{-9}${\cite{Bobeth2013}}\tabularnewline
              &  &  &  & $3.602\times10^{-9}${\cite{Shah2016}}\tabularnewline
            \cline{1-5} 
            \noalign{\smallskip}
 $B_{s}^{0}\rightarrow\tau^{+}\tau^{-}$ & $2.335\times10^{-13}$ & $3.361\times10^{-13}$ & $5.364\times10^{-7}$ & $7.22\times10^{-7}${\cite{Dimopoulos2011}}\tabularnewline
             &  &  &  & $(7.73\pm0.23)\times10^{-7}${\cite{Bobeth2013}}\tabularnewline
             &  &  &  & $7.647\times10^{-7}${\cite{Shah2016}}\tabularnewline
            \cline{1-5} 
           \noalign{\smallskip}
  $B_{s}^{0}\rightarrow e^{+}e^{-}$ & $2.577\times10^{-20}$ & $3.695\times10^{-20}$ & $5.921\times10^{-14}$ & $<2.8\times10^{-7}${\cite{PDGlatest}}\tabularnewline
              &  &  &  & $7.97\times10^{-14}${\cite{Dimopoulos2011}}\tabularnewline
              &  &  &  & $(8.54\pm0.55)\times10^{-14}${\cite{Bobeth2013}}\tabularnewline
              &  &  &  & $8.408\times10^{-14}${\cite{Shah2016}}\tabularnewline
 \hline 
 \end{tabular}
  \par\end{centering}
  \end{table}

     \begin{table}
      \begin{centering}
      \caption{\bf Branching ratio with corresponding radiative leptonic decay width for $B$ Meson. \label{tab:radiativelepto}}
       \begin{tabular}{cccccc}
      \hline
       \noalign{\smallskip}
   \multirow{2}{*}{Decay constant} & \multirow{2}{*}{$\Gamma(GeV)$} & \multicolumn{4}{c}{$BR$}\tabularnewline
    \cline{3-6} 
     \noalign{\smallskip}
    &  & This work & \bf \cite{Korchemsky2000}&\bf \cite{Yang2014}  & \bf \cite{Yang:2016} \tabularnewline
   \hline          
    \noalign{\smallskip}
       $fp$ & $1.51\times10^{-19}$ & $0.38\times10^{-6}$ &$0.23\times10^{-6}$& $1.66\times10^{-6}$ & $5.21\times10^{-6}$ \tabularnewline
             \noalign{\smallskip}
      $fpcor$ & $1.36\times10^{-19}$ & $0.34\times10^{-6}$ \tabularnewline
             \noalign{\smallskip}
               \hline 
       \end{tabular}
              \par\end{centering}
      \end{table}

      \begin{figure}
      \centering
      \includegraphics[bb=30bp 40bp 750bp 550bp,clip,width=0.72\textwidth]{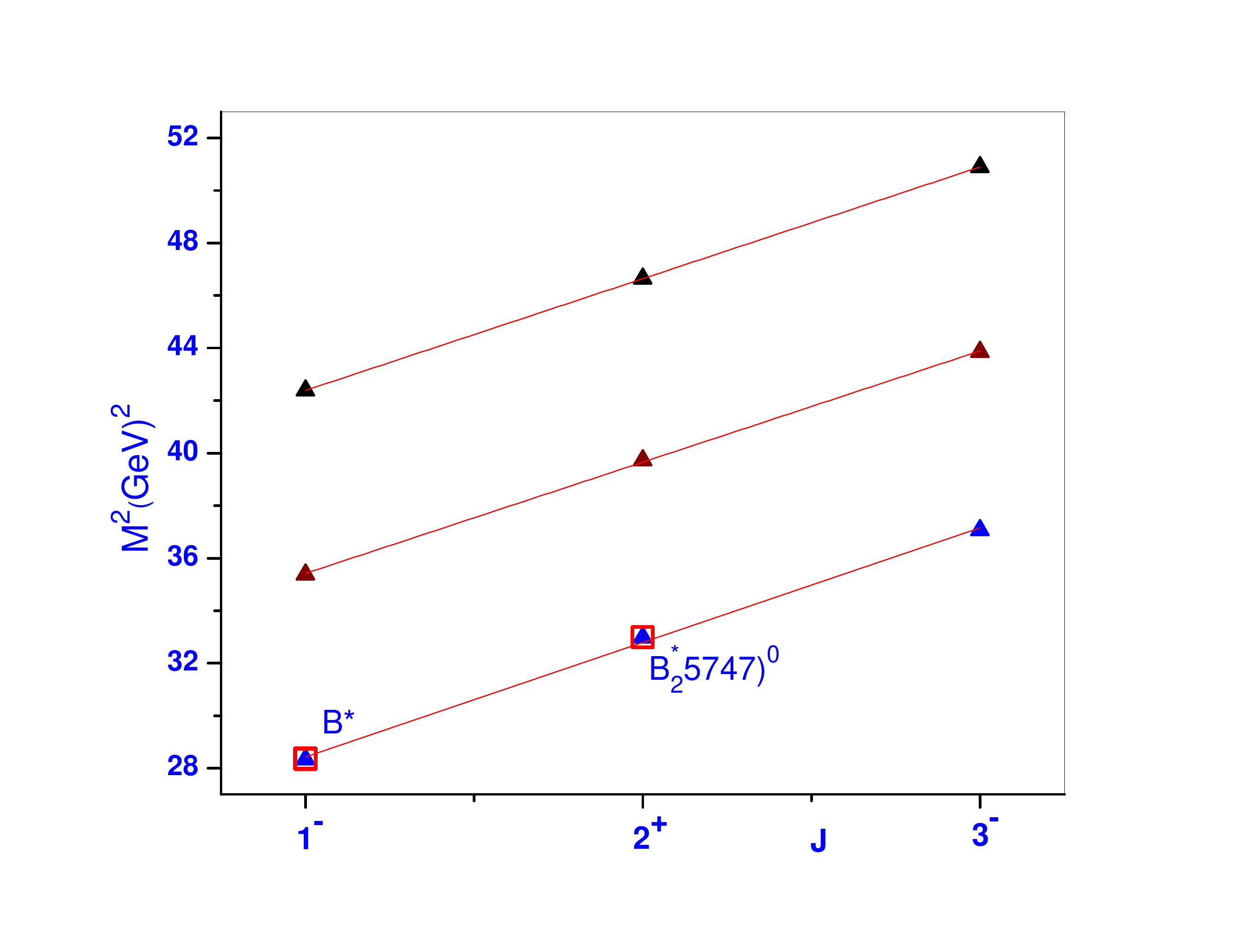}
      \caption{ Regge trajectory ($M^{2}\rightarrow J$) for the $B$ meson with natural parity.\label{fig:NPmesonB}}
      
      \end{figure}
      
      \begin{figure}
      \centering
      \includegraphics[bb=30bp 40bp 750bp 550bp,clip,width=0.72\textwidth]{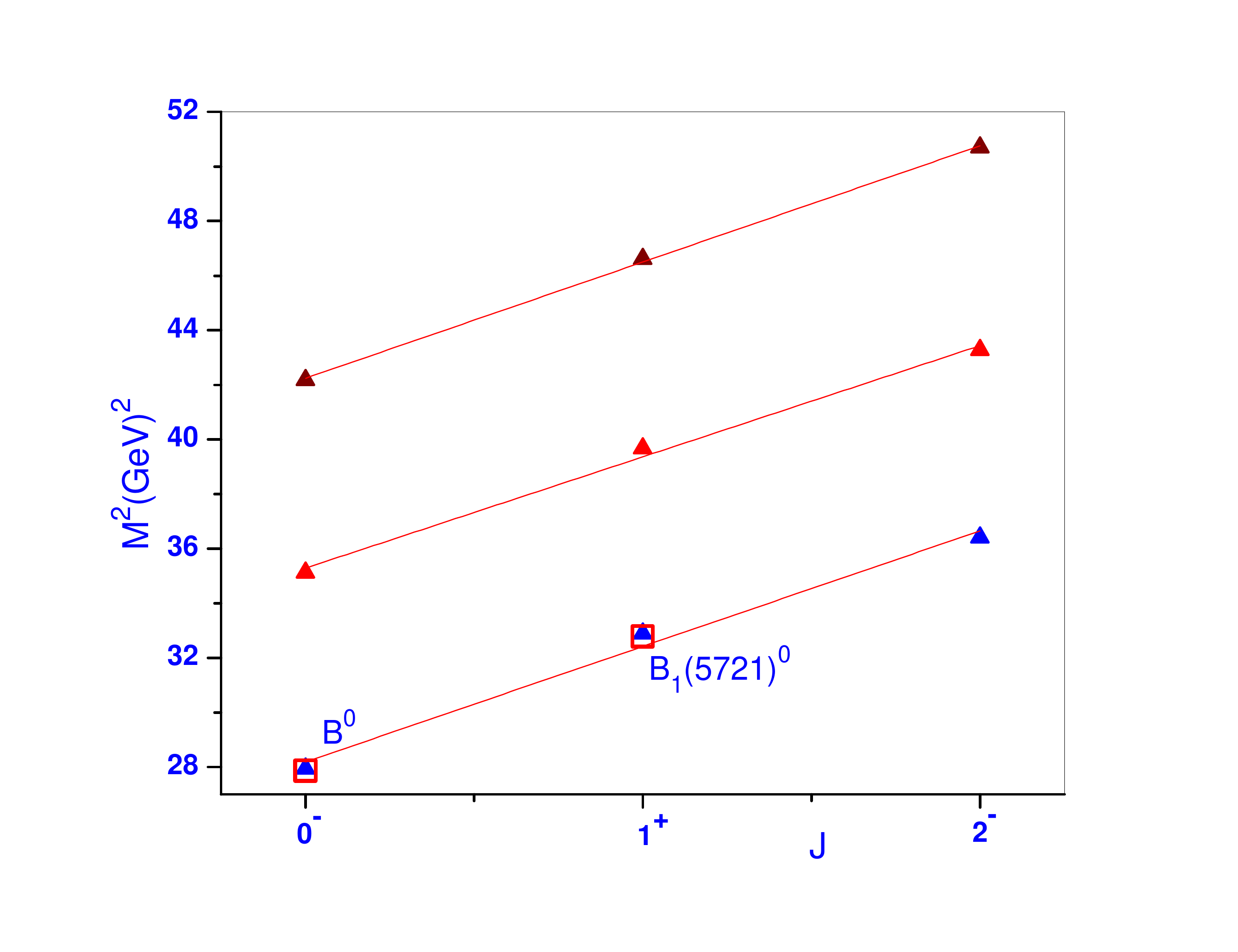}
      \caption{ Regge trajectory ($M^{2}\rightarrow J$) for the $B$ meson with unnatural parity. \label{fig:UNPmesonB}}
      \end{figure}

 \begin{multicols}{2}
 
  \section{Results and Discussion\label{sec:Resu}}
  
The center of weight masses for the S, P and D state are estimated using Equation(\ref{Eq:rai1-1}) and  Equation(\ref{Eq:rai2-1}) and results are tabulated in Table (\ref{tab:swavespinB}).{\bf In case of $B$ meson, the center of weight masses for the 1S, 2S, 3S, 1P, 2P, 1D and 2D state are in good agreement with outcome of other theoretical models, whereas masses for 4S, 5S, 3P and 3D are somewhat overestimated.  In case of $B_s$ meson, the center of weight masses are in good agreement with outcome of other theoretical models.} 
     
The estimated mass spectra for the $B$ and $B_S$ mesons are tabulated in Tables (\ref{tab:massesB}, \ref{tab:massesBs}) with the spectroscopic notation $n^{2S+1}L_{J}$. Mass spectra are also depicted graphically in figures (\ref{fig:MassB}) and (\ref{fig:MassBs}). The predicted masses of the $B$ and $B_s$ mesons are in close agreement to experimental observations. The difference of predicted and experimental observed value of mass is 7 MeV in $1^{1}S_{0}$,  2 MeV in $1^{3}S_{1}$,{\bf  20 MeV in $1^{3}P_{0}$, 7 MeV in $1P_{1}$ } and match in $1^{3}P_{2}$  state of the $B$ meson. Similarly, the difference of predicted and experimental observed value of mass is 1 MeV in  $1^{1}S_{0}$, 2 MeV in  $1^{3}S_{1}$, 1 MeV in  $1^{3}P_{1}$, 11 MeV in $1^{3}P'_{1}$ and is match with $1^{3}P_{2}$ state of the $B_s$ meson.

The predicted leptonic branching fractions for the $B$ meson are tabulated in Table (\ref{tab:leptobranch}). The predicted $BR_\mu$ and  $BR_e$ indicates that the predictions are in well agreement with the experimental outcomes, while $BR_\tau$  is slightly underestimated with respect to experimental observation. In the literature, various methods are used to calculate the radiative leptonic decay rate and branching ratio. In the Ref.\cite{Atwood1996}, calculated branching ratio ($B\rightarrow l\bar{\nu}{\gamma}$) of the order of $10^{-6}$ in a non-relativistic quark model. In Ref.\cite{Korchemsky2000} with perturbative QCD approach, it is found that the branching ratio of $B^{+}\rightarrow e^{+}\bar{\nu}{\gamma}$ is of the order of $10^{-6}$. In the factorization approach, it is found to be of order of $10^{-6}$ for the $B$ meson \cite{Yang2012,Yang2014,Yang2015,Yang:2016}. In the form factors parameterizing approach, it is found to be the order of $10^{-7}$ \cite{Korchemsky2000}. We also found branching ratio of the order of $10^{-7}$ for the $B$ meson.

Our predicted decay width and branching ratio of the rare leptonic decays $B_{s}\rightarrow l^{+}l^{-}$ and $B\rightarrow l^{+}l^{-},$$(l=\mu,\tau,e)$ are shown in the Tables (\ref{tab:rareleptoB} and \ref{tab:rareleptoBs}). The predicted branching ratio for $(B_{s}\rightarrow\mu^{+}\mu^{-})=2.529\times10^{-9}$
 and $(B\rightarrow\mu^{+}\mu^{-})=1.002\times10^{-10}$ are in excellent agreement with the experimental results published CMS and LHCb \cite{CMS:2015,Chatrchyan2013,Aaij2013}.   The predicted branching ratio for  $B_{s}^{0}\rightarrow e^{+}e^{-} = 5.921\times10^{-14}$  and $B^{0}\rightarrow e^{+}e^{-} = 2.345\times10^{-15}$ are in good agreement with Ref.\cite{Shah2016} (Dirac formalism),  Ref.\cite{Bobeth2013} (Standard model with reduced theoretical uncertainty) and Ref.\cite{Dimopoulos2011} (lattice results). 
 {\bf We have also predicted the branching } ratio with corresponding decay width of other rare leptonic $(l=\tau,e)$ decays for the $B$ and $B_{s}$ mesons. Due to the large uncertainty in experimental observations in rare leptonic $(l=\tau,e)$ decays for the $B$ and $B_{s}$ mesons, it is difficult to come-up with reasonable conclusion but outcomes we received are comparatively in good agreement with the prediction by other theoretical models.

\subsection{Regge trajectories \label{sec:reg}}
   
  The Regge trajectories play a vital role to identify any new (experimentally) excited state as well as to provide the information about quantum number of the particular state.  We use our predicted the ground, radial and orbital excited state masses of the $B$ and $Bs$ mesons, to constitute the Regge trajectories for the $(n,M^{2})$ and  $(J,M^{2})$ planes. Here,$M$ stands for mass of the $B$ and $Bs$ mesons, $n$ stands for the principal quantum number and $J$ is a total spin. 
  
\begin{figure}
     \centering
     \includegraphics[bb=30bp 40bp 750bp 550bp,clip,width=0.72\textwidth]{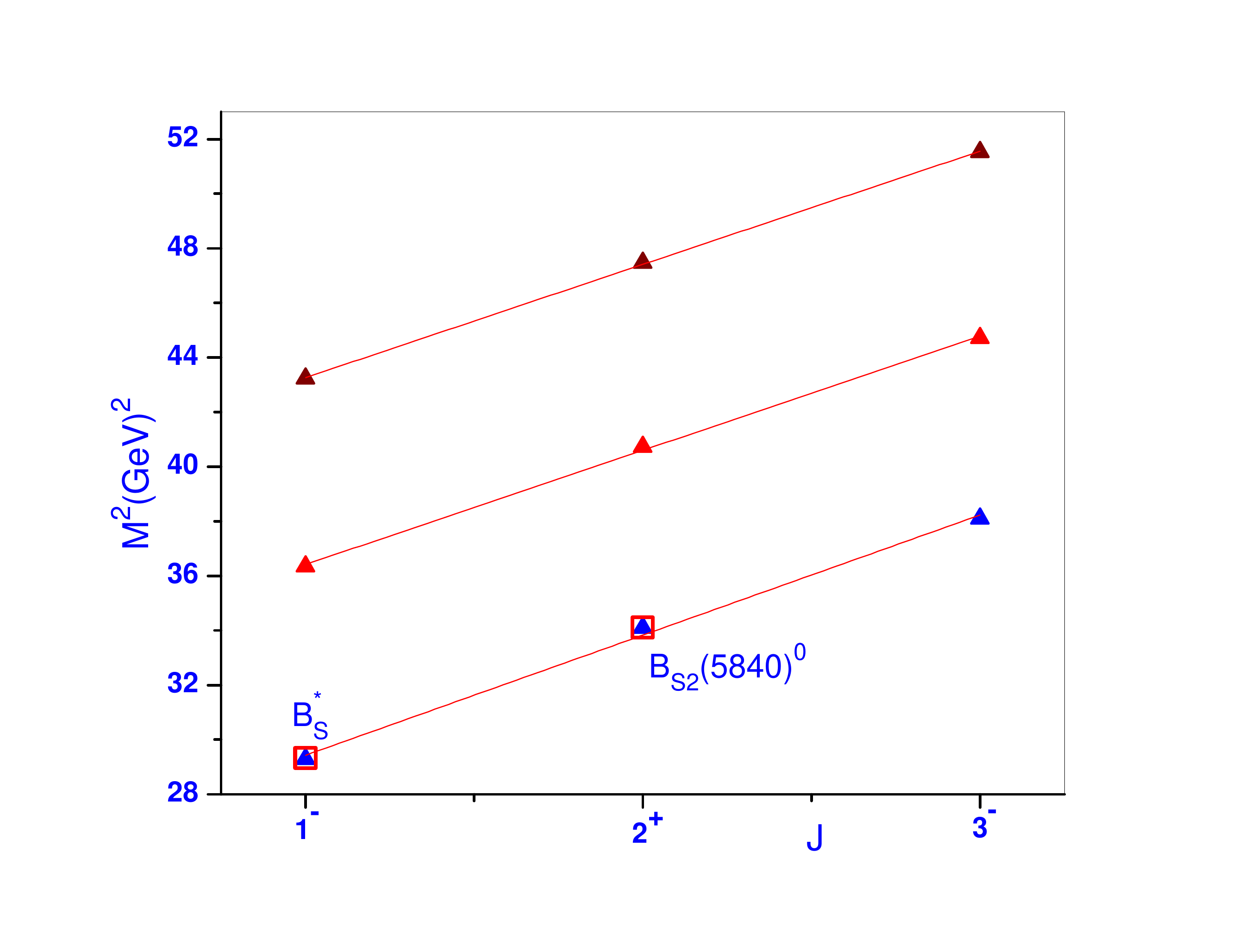}
     \caption{Regge trajectory ($M^{2}\rightarrow J$) for the $B_s$ meson with natural parity. \label{fig:NPmesonBs}}
     \end{figure}

\begin{figure}
     \centering
      \includegraphics[bb=30bp 40bp 750bp 550bp,clip,width=0.72\textwidth]{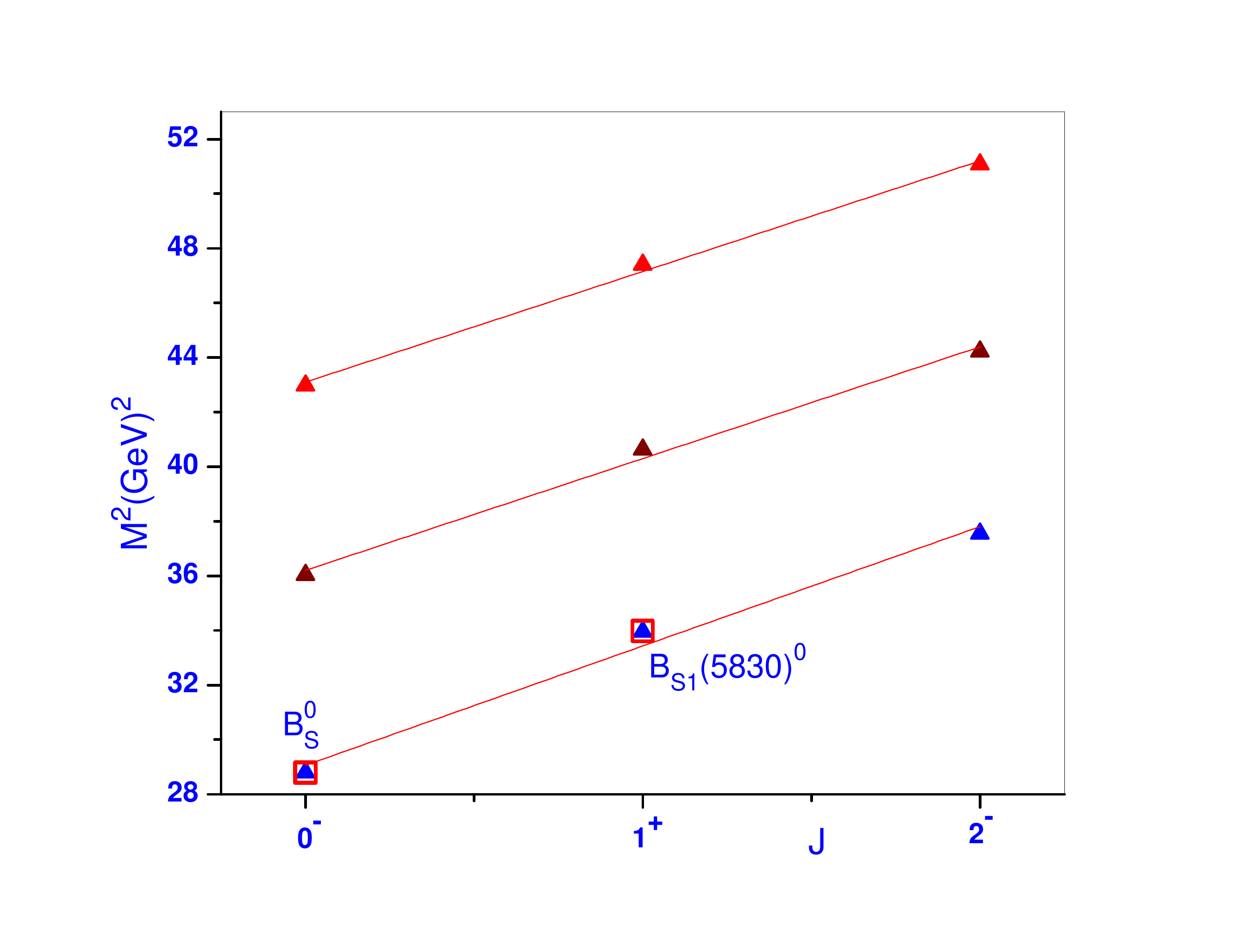}
     \caption{Regge trajectory ($M^{2}\rightarrow J$) for the $B_s$ meson with unnatural parity. \label{fig:UNPmesonBs}}
     \end{figure}

  The Regge trajectories with natural $(P=(-1)^{J}$; $J^{P}=1^{-},  2^{+}, 3^{-}$ ) and unnatural $(P=(-1)^{J-1}$; $J^{P}=0^{-}, 1^{+}, 2^{-}$ ) parity in the $(J,M^{2})$ plane, for the $B$ and $Bs$ mesons are depicted in figures (\ref{fig:NPmesonB}-\ref{fig:UNPmesonBs}). The masses predicted by our potential model are expressed by solid triangle and experimentally available values are presented by hollow square with the corresponding meson name. Straight $\chi^{2}$ fit lines were obtained for the predicted mass values. We have use following definition    
\begin{equation}
      J=\alpha M^{2}+\alpha_{0}\label{eq:J regge}
\end{equation}
   to find the slope ($\alpha$) and the intercept ($\alpha_{0}$). The slopes and intercepts for the $\chi^{2}$  fitted   $(J,M^{2})$ Regge trajectories are tabulated in Table(\ref{tab:alfa}).
    
\begin{figure}
     \centering
     \includegraphics[bb=30bp 40bp 750bp 550bp,clip,width=0.72\textwidth]{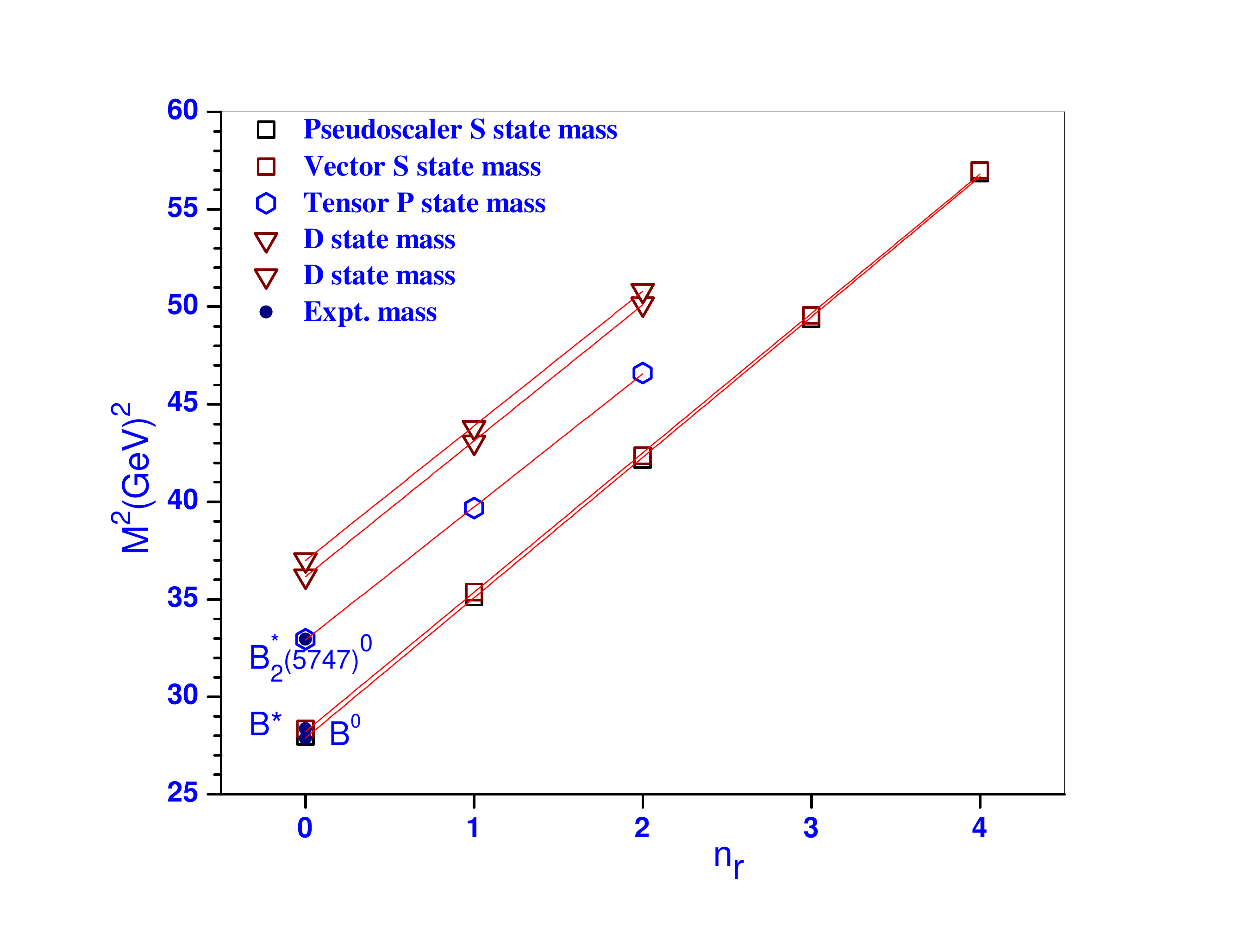}
     \caption{Regge trajectory ($M^{2}\rightarrow n_{r}$) for 
     the Pseudoscaler and vector $S$ state, excited $P$ and $D$ state masses of the $B$ meson.\label{fig:PsVmesonB}}
     \end{figure}

\begin{figure}
     \centering
     \includegraphics[bb=30bp 40bp 750bp 550bp,clip,width=0.72\textwidth]{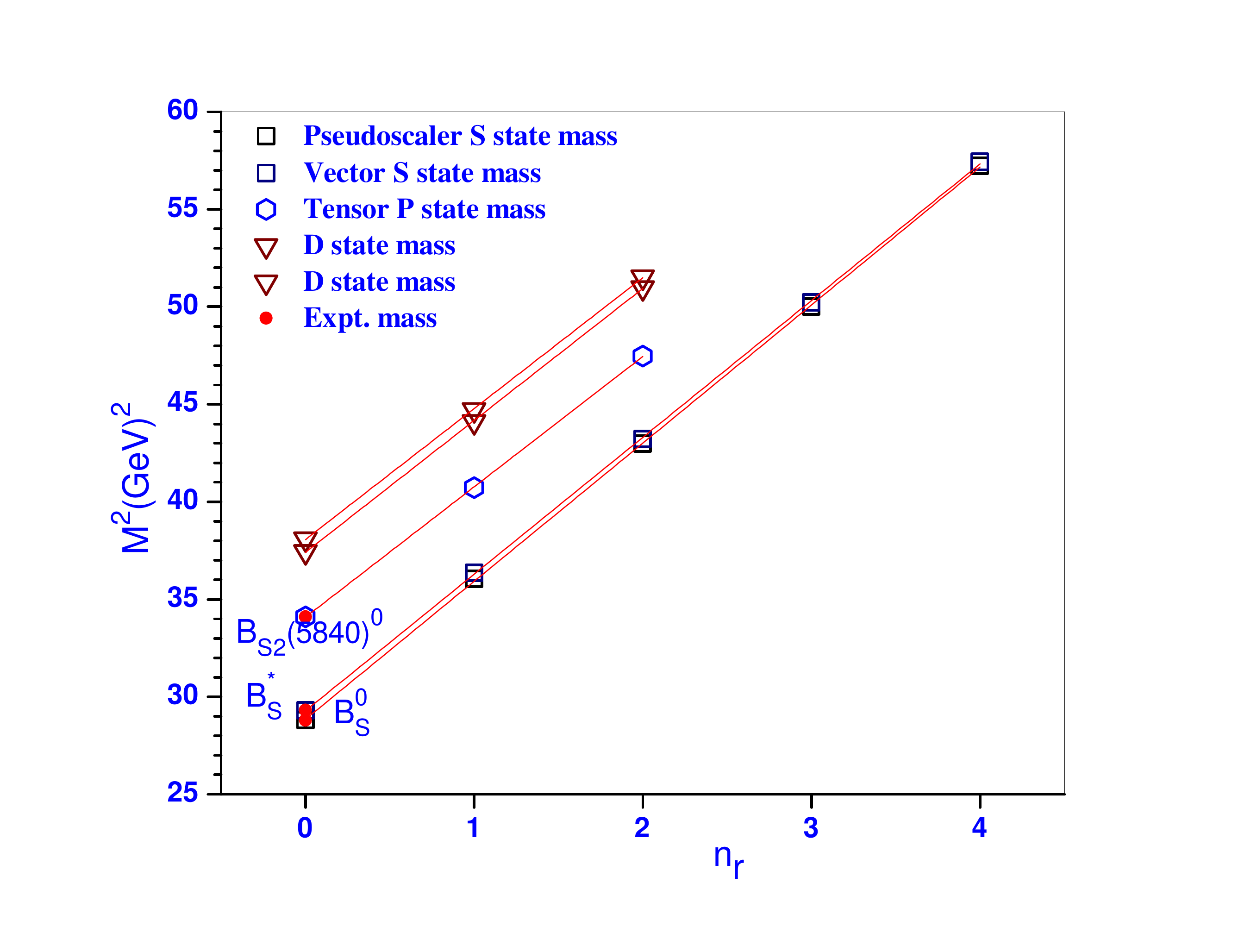}
     \caption{Regge trajectory ($M^{2}\rightarrow n_{r}$) for 
     the Pseudoscaler and vector $S$ state, excited $P$ and $D$ state masses of the $B_s$ meson.\label{fig:PsVmesonBs}}
     \end{figure}

\begin{figure}
     \centering
     \includegraphics[bb=30bp 40bp 750bp 550bp,clip,width=0.72\textwidth]{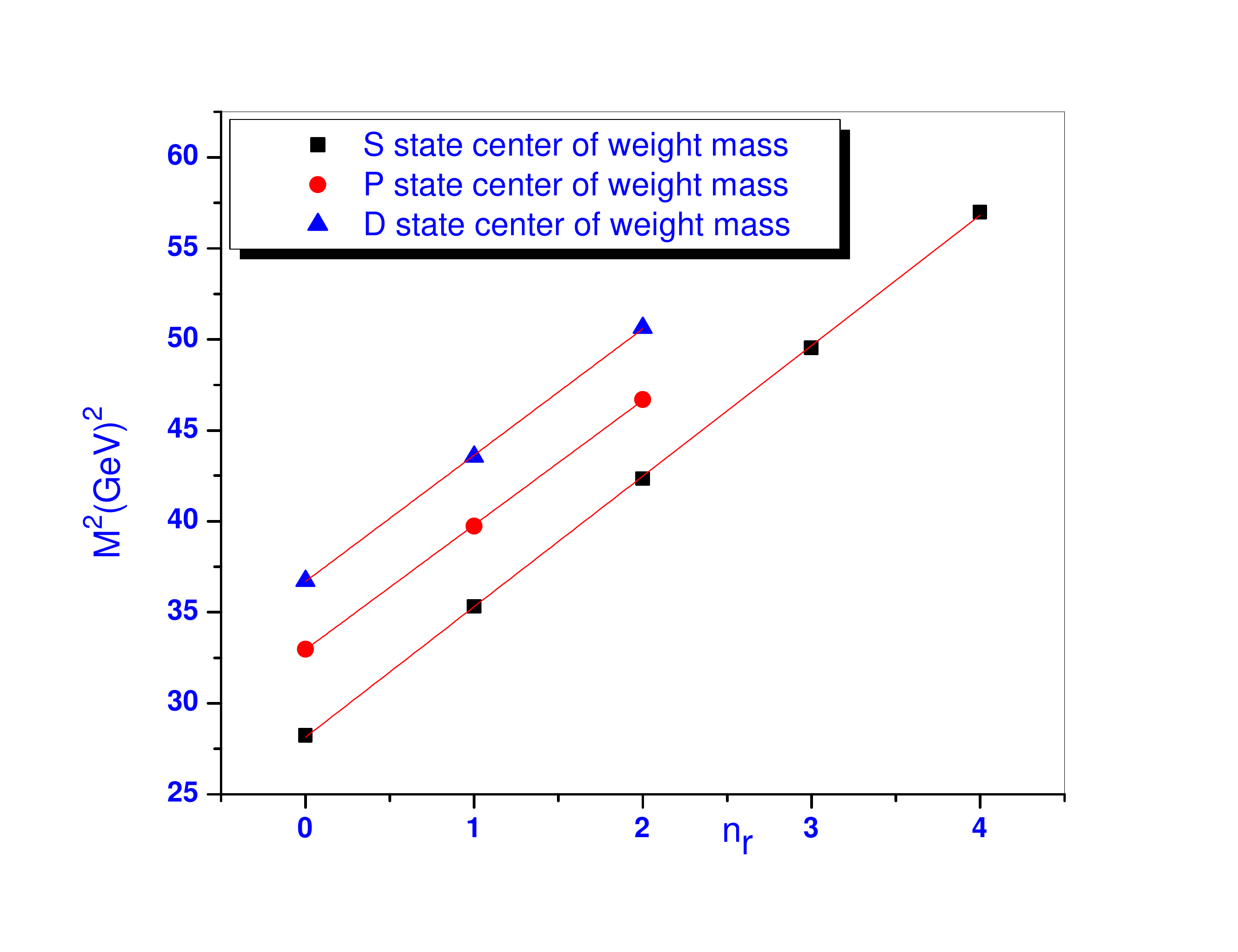}
     \caption{Regge trajectory ($M^{2}\rightarrow n_{r}$) for the S-P-D States center of weight mass for the $B$ meson.\label{fig:SavmesonB}}
     \end{figure}

\begin{figure}
     \centering
     \includegraphics[bb=30bp 40bp 750bp 550bp,clip,width=0.72\textwidth]{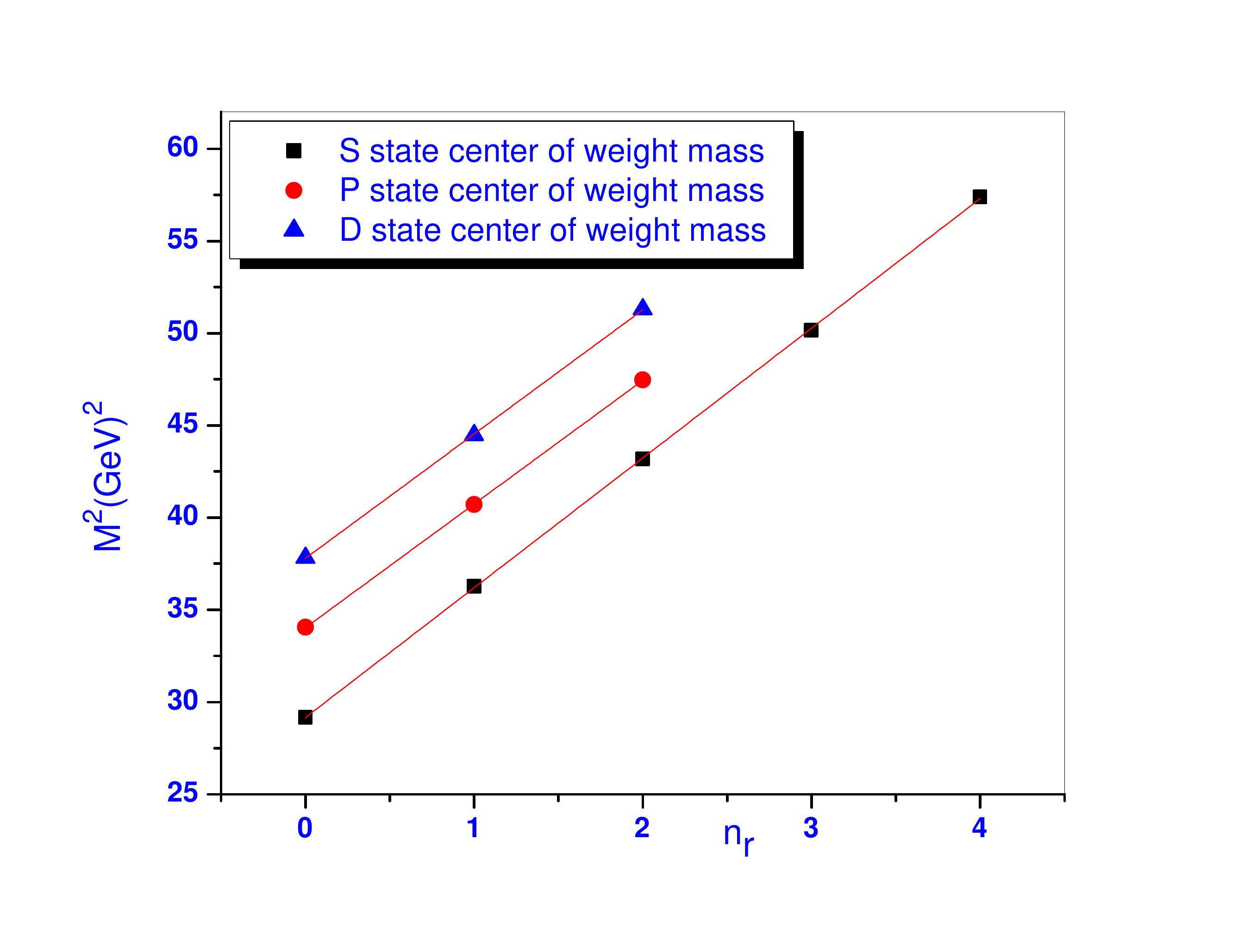}
     \caption{Regge trajectory ($M^{2}\rightarrow n_{r}$) for the S-P-D States center of weight mass of the $B_s$ meson.\label{fig:SavmesonBs}}
     \end{figure}

   Figures (\ref{fig:PsVmesonB},\ref{fig:PsVmesonBs}), depict the Regge trajectories using the Pseudoscaler ($J^{P}=0^{-}$) and vector ($J^{P}=1^{-}$) S state, excited P ($J^{P}=2^{+}$), D ($J^{P}=1^{-}$) and D ($J^{P}=3^{-}$)  state masses of the $B$ and $Bs$ mesons for $n_{r}= n-1$  principal quantum number in the $(n_{r},M^{2})$ plane. Available experimental values are given by solid dot with corresponding meson name. Figures (\ref{fig:SavmesonB},\ref{fig:SavmesonBs}), depict the Regge trajectories using the S, P and D  state center of weight masses of the $B$ and $Bs$ mesons for $n_{r}=(n-1)$  principal quantum number in the $(n_{r},M^{2})$ plane. We have use following definition
      \begin{equation}
      n_{r}=\beta M^{2}+\beta_{0}\label{eq:nr regge}
      \end{equation}
to find the slope ($\beta$) and the intercept ($\beta_{0}$). The slopes and intercepts for the $\chi^{2}$  fitted  $(n,M^{2})$ Regge trajectories are tabulated in Tables (\ref{tab:bita}, \ref{tab:Spinave}).
   
 With a comparison of the slopes, we found that the slope values $\alpha$ are larger than the slope value $\beta$ ones. The ratio of the mean of $\alpha$ and  $\beta$ is 1.68 and 1.62 for the $B$ and $B_S$ meson respectively. We observed that the estimated masses of  the $B$ and $B_s$ mesons are fit well to the linear trajectories in $(n,M^{2})$ and  $(J,M^{2})$ planes and are almost parallel to and equidistant with each other. With the help of Regge trajectories, we can identify any new (experimentally) excited states as well as ascertain the information about quantum number of the particular state. 
  
   \end{multicols}

   \begin{table}
    \caption{Fitted parameters of the $(J,\: M^{2})$ Regge trajectories with unnatural and natural parity.\label{tab:alfa} }
            \noindent \centering{}%
             \begin{tabular}{ccccc}
           \hline
             \noalign{\smallskip} 
           {Parity} & {Meson} & {Trajectory} & {$\alpha(GeV^{-2})$} & {$\alpha_{0}$}\tabularnewline
              \noalign{\smallskip}
           \hline 
              \noalign{\smallskip}
           \multirow{6}{*}{Unnatural} & \multirow{3}{*}{$B$$(b\bar{q})$} & Parent & $0.234\pm0.022$ & $-6.576\pm0.741$\tabularnewline
            &  & First daughter & $0.244\pm0.016$ & $-8.616\pm0.634$\tabularnewline
            &  & Second daughter & $0.235\pm0.005$ & $-9.813\pm0.261$\tabularnewline
           \cline{2-5} 
             \noalign{\smallskip}
            & \multirow{3}{*}{$B_{s}$$(b\bar{s})$} & Parent & $0.226\pm0.023$ & $-6.565\pm0.791$\tabularnewline
             &  & First daughter & $0.243\pm0.017$ & $-8.789\pm0.702$\tabularnewline
             &  & Second daughter & $0.246\pm0.013$ & $-10.605\pm0.611$\tabularnewline
           \hline 
             \noalign{\smallskip}
           \multirow{6}{*}{Natural } & \multirow{3}{*}{$B^{*}$$(c\bar{q})$} & Parent & $0.229\pm0.008$ & $-5.500\pm0.277$\tabularnewline
            &  & First daughter & $0.236\pm0.003$ & $-7.353\pm0.128$\tabularnewline
            &  & Second daughter & $0.235\pm0.0004$ & $-8.856\pm0.021$\tabularnewline
           \cline{2-5} 
              \noalign{\smallskip}
            & \multirow{3}{*}{ $B_{s}^{*}$$(c\bar{s})$} & Parent & $0.226\pm0.012$ & $-5.665\pm0.407$\tabularnewline
             &  & First daughter & $0.239\pm0.006$ & $-7.701\pm0.265$\tabularnewline
             &  & Second daughter & $0.241\pm0.003$ & $-9.425\pm0.149$\tabularnewline
           \hline
           \end{tabular}
           \end{table}

   \begin{table}
   \caption{Fitted slope and intercept for the $(n_{r},\: M^{2})$ Regge trajectories. \label{tab:bita}}
   
   \noindent \begin{centering}
   \begin{tabular}{ccccc}
   \hline 
      \noalign{\smallskip}
   Meson & State & $J^P$ & $\beta(GeV^{-2})$ & $\beta_{0}$\tabularnewline
   \hline 
    \noalign{\smallskip}
   \multirow{5}{*}{$B$}  & S &$0^{-}$  & $0.139\pm0.0008$ & $-3.873\pm0.037$\tabularnewline
   &S & $1^{-}$ & $0.139\pm0.001$ & $-3.946\pm0.045$\tabularnewline
    &P &$2^{+}$  & $0.146\pm0.001$ & $-4.822\pm0.047$\tabularnewline
   &D & $1^{-}$ & $0.143\pm0.001$ & $-5.182\pm0.057$\tabularnewline
   &D &$3^{-}$  & $0.144\pm0.001$ & $-5.357\pm0.067$\tabularnewline
  
   \hline 
     \noalign{\smallskip}
    
   \multirow{5}{*}{$B_{s}$}  &S &$0^{-}$  & $0.141\pm0.0006$ & $-4.071\pm0.026$\tabularnewline
    & S&$1^{-}$  & $0.142\pm0.0006$ & $-4.173\pm0.028$\tabularnewline
     &P &$2^{+}$  & $0.149\pm0.0007$ & $-5.097\pm0.030$\tabularnewline
    &D &$1^{-}$  & $0.147\pm0.001$ & $-5.526\pm0.051$\tabularnewline
    &D &$3^{-}$  & $0.149\pm0.001$ & $-5.669\pm0.057$\tabularnewline
   \hline 
  
   \end{tabular}
   \par\end{centering}
   
   \end{table}

\begin{table}
  \caption{Fitted parameters of Regge trajectory $(n_{r},\: M^{2})$ for the S-P-D states center of weight mass.\label{tab:Spinave} }
      \noindent \centering{}%
     \begin{tabular}{cccc}
     \hline 
       \noalign{\smallskip}
     Meson & Trajectory & $\beta(GeV^{-2})$ & $\beta_{0}$\tabularnewline
     \hline 
       \noalign{\smallskip}
     \multirow{3}{*}{$B$$(b\bar{q})$} & S Satate & $0.139\pm0.001$ & $-3.922\pm0.042$\tabularnewline
      & P State & $0.145\pm0.001$ & $-4.810\pm0.042$\tabularnewline
      & D State & $0.143\pm0.001$ & $-5.283\pm0.061$\tabularnewline
     \hline 
       \noalign{\smallskip}
      \multirow{3}{*}{$B_{s}$$(b\bar{s})$} & S Satate & $0.142\pm0.0006$ & $-4.147\pm0.027$\tabularnewline
      & P State & $0.149\pm0.0006$ & $-5.075\pm0.024$\tabularnewline
      & D State & $0.148\pm0.0012$ & $-5.609\pm0.054$\tabularnewline
     
     \hline
     \end{tabular}
     \end{table}

 \begin{multicols}{2}

\section{Conclusion\label{sec:conclusion}}
{\bf Our predicted masses of the $S-P-D$ states of the $B$ and $B_s$ mesons are very close to the available experimental values. The excited states($4S$, $5S$, $3P$ and $3D$) masses are higher than other theoretical estimates. The estimated relativistic correction to the kinetic energy term, which is found less than $1\%$, while that of to the potential energy term is found to be around $(3-5)\%$.} With limited experimental observations of excited states of the $B$ and $B_s$ mesons, the LHCb collaborator's recent measurements show the possibility to update our knowledge of the excited states of these mesons. For the $B$ meson, 1P states $B_{1}(5721)$ and $B_{2}^{\star}(5747)$ are in good agreement with our prediction \cite{PDGlatest,Aaij:2015}. { \bf We have assigned newly observed states B(5970)\cite{Aaltonen:2013} to $2^3S_1$ state of the $B$ meson.} For the $B_s$ meson, 1P experimental states $B_{S1}^{\star}(5830)$  and $B_{S2}^{\star}(5840)$ are in excellent agreement with our predicted values. These are the few excited states measurements, but in new future LHCb will provide the new results in bottom meson spectroscopy. 

The Regge trajectories for the $B$ and $B_s$ meson, in both ($J, M^2$)  and ($n_r, M^2$) plane are almost linear, equidistant and parallel. Regge trajectories provide the information about quantum numbers of the particular state and useful to identify any new excited state (experimentally). From  figures (\ref{fig:NPmesonB}-\ref{fig:UNPmesonBs}), we found that experimental states are nicely sitting on due straight lines without deviation.

The pseudoscaler decay constants with QCD correction for both mesons are underestimated compare to other theoretical models estimates.  Our calculated leptonic branching fractions of the $B$ meson  are fairly closed to the PDG\cite{PDGlatest} values. Our calculated  radiative leptonic branching ratio of the order of $10^{-6}$ is in good agreement with branching ratio obtained with form factor parameterizing approach Ref.\cite{Korchemsky2000}. {\bf The estimated rare leptonic decay width as well as branching ratio of the $B$ and $B_s$ mesons are compared with  experimental observations  and other theoretical estimates \cite{Shah2016,Chatrchyan2013,Aaij2013,Dimopoulos2011},  results are in good agreement (see Tables (\ref{tab:rareleptoB},\ref{tab:rareleptoBs}))}. 

Finally, this study may help the current experimental facilities to identify the new states as well as their $J^{PC}$ values of the $B$ and $B_s$ mesons.\\

{\bf Acknowledgments}

A. K. Rai acknowledge the financial support extended by Department of Science of Technology, India  under SERB fast track scheme SR/FTP /PS-152/2012.\\

\bibliographystyle{epj}
\bibliography{myref}

\begin{thebibliography}{53}

\bibitem{PDGlatest}
C.P. et~al.(Particle Data~Group), Chinese Physics C \textbf{40}, 100001 (2016)

\bibitem{Lu:2016}
Q.F. Lu, T.T. Pan, Y.Y. Wang, E.~Wang, D.M. Li, Phys. Rev. \textbf{D94}(7),
  074012 (2016)

\bibitem{Shah2016}
M.~Shah, B.~Patel, P.~Vinodkumar, Phys. Rev. \textbf{D93}(9), 094028 (2016)

\bibitem{Liu2016}
J.B. Liu, C.D. Lu, arXiv:1605.05550 [hep-ph]  (2016)

\bibitem{Liu2015}
J.B. Liu, M.Z. Yang, Chin. Phys. \textbf{C40}(7), 073101 (2016)

\bibitem{Aaij:2015}
R.~Aaij et~al., JHEP \textbf{04}, 024 (2015)

\bibitem{chen:2016}
H.X. Chen, W.~Chen, X.~Liu, Y.R. Liu, S.L. Zhu, Rept. Prog. Phys. \textbf{80},
  076201 (2017), \texttt{1609.08928}

\bibitem{Godfrey1985}
S.~Godfrey, N.~Isgur, Phys. Rev. D \textbf{32}, 189 (1985)

\bibitem{Colangelo1993}
P.~Colangelo, F.~De~Fazio, G.~Nardulli, Phys. Lett. \textbf{B316}, 555 (1993)

\bibitem{DiPierro2001}
M.~Di~Pierro, E.~Eichten, Phys. Rev. \textbf{D64}, 114004 (2001)

\bibitem{Yang2012}
M.Z. Yang, Eur. Phys. J. \textbf{C72}, 1880 (2012)

\bibitem{Ebert2011jc}
D.~Ebert, R.~Faustov, V.~Galkin, Eur. Phys. J. \textbf{C71}, 1825 (2011)

\bibitem{Eisner2013}
A.M. Eisner, \emph{{Recent Results on Radiative and Electroweak Penguin Decays
  of B Mesons at BaBar}}, in \emph{{(DPF 2013), USA, August 13-17, 2013}}
  (2013)

\bibitem{Ebert2010}
D.~Ebert, R.~Faustov, V.~Galkin, Eur. Phys. J. \textbf{C66}, 197 (2010)

\bibitem{Yang:2012}
J.C. Yang, M.Z. Yang, Mod. Phys. Lett. \textbf{A27}, 1250120 (2012)

\bibitem{Gupta1995}
S.N. Gupta, J.M. Johnson, Phys. Rev. D \textbf{51}(1), 168 (1995)

\bibitem{Hwang1997}
D.S. Hwang, C.~Kim, W.~Namgung, Phys. Lett. \textbf{B406}, 117 (1997)

\bibitem{Kher:2017}
V.~Kher, N.~Devlani, A.K. Rai (2017), \texttt{1704.00439}

\bibitem{Koma2006}
Y.~Koma, M.~Koma, H.~Wittig, Phys. Rev. Lett \textbf{97}, 122003 (2006)

\bibitem{Eichten1978}
E.~Eichten, K.~Gottfried, T.~Kinoshita, K.D. Lane, T.M. Yan, Phys. Rev. D
  \textbf{17}(11), 3090 (1978)

\bibitem{Devlani2013}
N.~Devlani, A.K. Rai, Int. J. Theor. Phys. \textbf{52}, 2196 (2013), ISSN
  0020-7748

\bibitem{Devlani2011}
N.~Devlani, A.K. Rai, Phys. Rev. D \textbf{84}, 074030 (2011)

\bibitem{Rai2008}
A.K. Rai, B.~Patel, P.C. Vinodkumar, Phys. Rev. C \textbf{78}(5), 055202 (2008)

\bibitem{Rai2002}
A.K. Rai, R.H. Parmar, P.C. Vinodkumar, J. Phys. G: Nucl. Part. Phys.
  \textbf{28}(8), 2275 (2002)

\bibitem{Eichten1994}
E.J. Eichten, C.~Quigg, Phys. Rev. D \textbf{49}(11), 5845 (1994)

\bibitem{Gromes1984}
D.~Gromes, Z. Phys. \textbf{C26}, 401 (1984)

\bibitem{Gershtein1995}
S.~Gershtein, V.~Kiselev, A.~Likhoded, A.~Tkabladze, Phys. Usp. \textbf{38}, 1
  (1995)

\bibitem{Godfrey:2016}
S.~Godfrey, K.~Moats, E.S. Swanson, Phys. Rev. \textbf{D94}(5), 054025 (2016)

\bibitem{Barnes:2002}
T.~Barnes, N.~Black, P.R. Page, Phys. Rev. \textbf{D68}, 054014 (2003)

\bibitem{Silverman1988}
D.~Silverman, H.~Yao, Phys. Rev. D \textbf{38}(1), 214 (1988)

\bibitem{Cai-Dian2003}
C.D. Lu, G.L. Song, Phys. Lett. \textbf{B562}, 75 (2003)

\bibitem{Bobeth2013}
C.~Bobeth, M.~Gorbahn, T.~Hermann, M.~Misiak, E.~Stamou, M.~Steinhauser, Phys.
  Rev. Lett. \textbf{112}, 101801 (2014)

\bibitem{Bobeth2013a}
C.~Bobeth, M.~Gorbahn, E.~Stamou, Phys. Rev. \textbf{D89}(3), 034023 (2014)

\bibitem{Buchalla1993}
G.~Buchalla, A.J. Buras, Nucl. Phys. \textbf{B400}, 225 (1993)

\bibitem{Buras1998}
A.J. Buras, in \emph{{Weak Hamiltonian, CP violation and rare decays,
  Proceedings, France, July 28-September 5, 1997. Pt. 1, 2}} (1998), pp.
  281--539

\bibitem{Lee2015}
H.S. Lee, \emph{{sin$^2 \theta_W$ theory and new physics}}, in \emph{{10th
  International Workshop on e+e- collisions from Phi to Psi (PHIPSI15), China,
  September 23-26, 2015}} (2015)

\bibitem{VanRoyen1967}
R.~Van~Royen, V.~Weisskopf, Nuovo Cim. \textbf{A50}, 617 (1967)

\bibitem{Braaten1995}
E.~Braaten, S.~Fleming, Phys. Rev. D \textbf{52}(1), 181 (1995)

\bibitem{Sun:2014}
Y.~Sun, Q.T. Song, D.Y. Chen, X.~Liu, S.L. Zhu, Phys. Rev. \textbf{D89}(5),
  054026 (2014)

\bibitem{Devlani2012}
N.~Devlani, A.~Rai, Eur. Phys. J. \textbf{A48}, 104 (2012)

\bibitem{Lahde2000}
T.~Lahde, C.~Nyfalt, D.~Riska, Nucl. Phys. \textbf{A674}, 141 (2000)

\bibitem{Aaltonen:2013}
T.A. Aaltonen et~al., Phys. Rev. \textbf{D90}(1), 012013 (2014)

\bibitem{Affolder:2001}
T.~Affolder et~al. (CDF), Phys. Rev. \textbf{D64}, 072002 (2001)

\bibitem{Chatrchyan2013}
S.~Chatrchyan et~al., Phys. Rev. Lett. \textbf{111}, 101804 (2013)

\bibitem{Aaij2012}
R.~Aaij et~al. (LHCb), Phys. Rev. Lett. \textbf{110}(2), 021801 (2013)

\bibitem{Aaij2013}
R.~Aaij et~al. (LHCb), Phys. Rev. Lett. \textbf{111}, 101805 (2013)

\bibitem{Dimopoulos2011}
P.~Dimopoulos et~al. (ETM), JHEP \textbf{01}, 046 (2012)

\bibitem{Korchemsky2000}
G.P. Korchemsky, D.~Pirjol, T.M. Yan, Phys. Rev. \textbf{D61}, 114510 (2000)

\bibitem{Yang2014}
J.C. Yang, M.Z. Yang, Nucl. Phys. \textbf{B889}, 778 (2014)

\bibitem{Yang:2016}
J.C. Yang, M.Z. Yang, Nucl. Phys. \textbf{B914}, 301 (2017)

\bibitem{Atwood1996}
D.~Atwood, G.~Eilam, A.~Soni, Mod. Phys. Lett. \textbf{A11}, 1061 (1996)

\bibitem{Yang2015}
J.C. Yang, M.Z. Yang, Mod. Phys. Lett. \textbf{A31}(03), 1650012 (2015)

\bibitem{CMS:2015}
V.~Khachatryan et~al., Nature \textbf{522}, 68 (2015)

\end{thebibliography}

\end{multicols}

\vspace{-1mm}
\centerline{\rule{80mm}{0.1pt}}
\vspace{2mm}

\begin{multicols}{2}

\end{multicols}

\clearpage

\end{document}